\documentclass[preprint,review,12pt,authoryear]{elsarticle}

\usepackage{subcaption}
\usepackage{amssymb}
\usepackage{amsmath}
\usepackage{booktabs}
\usepackage{graphicx}
\usepackage{hyperref}
\usepackage[labelfont=bf,labelsep=period]{caption}

\journal{ENG APPL COMP FLUID}

\begin{document}

\begin{frontmatter}

\title{DRLinSPH: An open-source platform using deep reinforcement learning and SPHinXsys for fluid-structure-interaction problems}

\author[inst1]{Mai Ye}

\author[inst2]{Hao Ma}

\author[inst3]{Yaru Ren}

\author[inst4]{Chi Zhang}

\author[inst1]{Oskar J. Haidn}

\author[inst1]{Xiangyu Hu\texorpdfstring{\corref{cor1}}{}}
\ead{xiangyu.hu@tum.de}

\cortext[cor1]{Corresponding author}

\affiliation[inst1]{organization={TUM School of Engineering and Design, Technical University of Munich},
            postcode={85748}, 
            city={Garching},
            country={Germany}}

\affiliation[inst2]{organization={School of Aerospace Engineering, Zhengzhou University of Aeronautics}, 
            postcode={450046}, 
            city={Zhengzhou},
            country={China}}

\affiliation[inst3]{organization={State Key Laboratory of Hydraulics and Mountain River Engineering, Sichuan University}, 
            postcode={610065}, 
            city={Chengdu},
            country={China}}

\affiliation[inst4]{organization={Huawei Technologies Munich Research Center}, 
            postcode={80992}, 
            city={Munich},
            country={Germany}}

\begin{abstract}
Fluid-structure interaction (FSI) problems are characterized by strong nonlinearities arising from complex interactions between fluids and structures. These pose significant challenges for traditional control strategies in optimizing structural motion, often leading to suboptimal performance. In contrast, deep reinforcement learning (DRL), through agent interactions within numerical simulation environments and the approximation of control policies using deep neural networks (DNNs), has shown considerable promise in addressing high-dimensional FSI problems. Additionally, smoothed particle hydrodynamics (SPH) offers a flexible and efficient computational approach for modeling large deformations, fractures, and complex interface movements inherent in FSI, outperforming traditional grid-based methods. In this work, we present DRLinSPH, an open-source Python platform that integrates the SPH-based numerical environment provided by the open-source software SPHinXsys with the mature DRL platform Tianshou to enable parallel training for FSI problems. DRLinSPH has been successfully applied to four FSI scenarios: sloshing suppression using rigid and elastic baffles, optimization of wave energy capture through an oscillating wave surge converter (OWSC), and muscle-driven fish swimming in vortices. The results demonstrate the platform's accuracy, stability, and scalability, highlighting its potential to advance industrial solutions for complex FSI challenges.
\end{abstract}


\end{frontmatter}

\section{Introduction}
Reinforcement learning (RL) is a fundamental method in machine learning. The main idea is to learn the best decision-making policies through trial and error by continuously interacting with an environment \citep{sutton2018reinforcement}. Since the 1980s, RL has developed from fundamental ideas like the Markov decision process (MDP) and dynamic programming (DP) \citep{bellman1957markovian, bryson1996optimal}. Important algorithms like temporal difference (TD) learning and Q-learning have greatly improved RL's theoretical structure \citep{sutton1981toward, watkins1992q}, allowing it to perform well in simple, low-dimensional spaces. In recent years, the fast growth of deep neural networks (DNNs) has led to the rise of deep reinforcement learning (DRL) \citep{sze2017efficient, arulkumaran2017deep}. Unlike traditional RL, DRL uses the powerful abilities of DNNs to extract features and represent data, enabling it to directly learn the mapping from state to action in high-dimensional continuous spaces, making it capable of handling complex, nonlinear problems. This progress has made DRL widely used in various areas, especially in robot control \citep{ibarz2021train}, natural language processing \citep{he2015deep}, complex games \citep{silver2017mastering}, and autonomous vehicle navigation \citep{aradi2020survey}.

Over the past few years, DRL has already found applications in fluid mechanics and mechanical engineering, primarily in structural optimization and active flow control (AFC) \citep{vignon2023recent}. Within the former, traditional optimization algorithms have succeeded in numerous practical applications yet possess certain limitations \citep{kenway2016multipoint}. For instance, gradient-based methods exhibit sensitivity to the initial starting point in strongly nonlinear problems, often leading to instability and a tendency to become trapped in local optima \citep{skinner2018state}. On the other hand, non-gradient algorithms, such as genetic algorithms, typically demand significant computational resources \citep{yamazaki2008aerodynamic}, while particle swarm optimization struggles to impose practical constraints on design parameters \citep{hassan2005comparison}. DRL offers an alternative approach for solving nonlinear and non-convex optimization problems \citep{viquerat2021direct}. Training an optimized structure within a limited time frame is possible by leveraging appropriately designed reward functions and exploring policies without relying on prior experience. \cite{viquerat2021direct} were the first to apply the proximal policy optimization (PPO) algorithm in DRL for direct airfoil shape optimization, achieving an optimal airfoil based on a reward function maximizing the lift-to-drag ratio. \cite{keramati2022deep} also utilized the PPO algorithm to optimize the thermal shape of a 2D heat exchanger based on Bézier curves. \cite{ma2024comprehensive} employed the deep Q-network (DQN) algorithm for the structural optimization of rocket engine nozzles. Notably, they were the first to integrate a U-Net-based vision transformer (ViT) and convolutional neural network (CNN) as surrogate models for computational fluid dynamics (CFD) within the RL environment, significantly reducing training time.

On the other hand, applications of DRL in AFC primarily encompass areas such as microfluidics, heat transfer, drag reduction, sloshing suppression, and swimming \citep{viquerat2022review}. \cite{dressler2018reinforcement} employed the DQN algorithm to adjust the laminar flow interface between two fluids in microfluidic channels at low Reynolds numbers and control the droplet size by modulating flow rates. \cite{lee2019case} applied DRL to optimize flow sculpting in microfluidic devices and compared it with traditional genetic algorithms. Their results indicated that DRL was more efficient in achieving the objectives and demonstrated a certain level of transferability. \cite{hachem2021deep} effectively optimized the enhancement of heat transfer in a two-dimensional (2D) and three-dimensional (3D) cavity with uneven wall temperature distributions using a degenerate version of the PPO algorithm and also mitigated the wall temperature non-uniformity caused by impingement cooling. In drag reduction, \cite{rabault2019artificial} were the first to control the Kármán vortex street in a 2D cylinder flow at a moderate Reynolds number ($Re = 100$). Using the PPO algorithm to regulate the flow from two jets positioned above and below the cylinder, they achieved an 8\% reduction in the drag coefficient while maintaining a nearly constant lift coefficient. Building on their work, \cite{ren2021applying}demonstrated that DRL agents can still discover effective control strategies at a higher Reynolds number ($Re = 1000$). \cite{paris2021robust} introduced a new algorithm, S-PPO-CMA, to optimize sensor placement and reduce the number of sensors while maintaining the performance of the DRL agent, achieving an 18.4\% reduction in drag. \cite{han2022deep} mentioned that as the Reynolds number increases, the drag reduction effect of active control on the cylinder becomes more pronounced. \cite{mao2022active} introduced an MDP with time delays and increased the number of jets, reducing the magnitude of drag and lift fluctuations by approximately 90\%. \cite{wang2022drlinfluids} developed a platform DRLinFluids based on OpenFOAM and the DRL framework Tianshou and applied the soft actor-critic (SAC) algorithm for active flow control on a square cylinder, achieving a drag reduction of approximately 13.7\%. Additionally, \cite{wang2022deep} optimized the NACA 0012 airfoil, achieving a 27.0\% reduction in drag and a 27.7\% increase in lift. \cite{ren2024adaptive} used DRL to effectively control transonic buffet (unstable flow) and transonic buffeting (structural vibration) in nonlinear fluid-structure interaction (FSI) systems. \cite{wang2023deep} utilized transfer learning to apply a DRL agent trained at low Reynolds numbers to train bluff body flows at high Reynolds numbers, significantly reducing the training time. 

Notably, the research on sloshing suppression and swimming differs from the others in that it requires consideration of FSI problems. \cite{xie2021sloshing} employed the twin delayed deep deterministic policy gradient (TD3) algorithm with behavior cloning to actively control two baffles under 2D sloshing in a tank, achieving an 81.48\% reduction in sloshing. Subsequently, \cite{xie2022active} applied active control to breakwaters under different, longer wave periods, demonstrating that the wave dissipation performance was superior to that of passive breakwaters. \cite{verma2018efficient} developed a DRL algorithm based on deep recurrent neural networks (RNNs) that accurately captures the interaction between fish and the vortices in schooling. The trained agent fish can utilize the shed vortices from the leading swimmer's wake to enhance propulsion efficiency, achieving energy savings. \cite{gunnarson2021learning} used DRL to explore the optimal path for a swimmer moving at a fixed speed through a 2D vortex field, finding that perceiving the velocity field significantly aids in training the agent. \cite{zhu2021numerical} combined the deep recurrent Q-network (DRQN) with the lattice Boltzmann method (LBM) to train fish on locating specific targets in still water and maintaining stable swimming in a Kármán vortex street. \cite{wang2024learn} utilized the PPO algorithm combined with a transformer architecture to optimize the motion trajectory of a NACA0016 flap, resulting in significant increases in thrust and efficiency compared to sinusoidal motion. \cite{cui2024enhancing} also used DRL to maximize propulsion efficiency and minimize energy consumption in a bio-mimetic robotic fish.

The examples above demonstrate the successful applications of DRL in fluid mechanics. Notably, current CFD environments combined with DRL are primarily based on grid-based methods, such as the finite element method (FEM) \citep{tezduyar1992new} and the immersed boundary method (IBM) \citep{peskin2002immersed}, utilizing custom PDE solvers or open-source platforms like OpenFOAM to solve the Navier-Stokes (NS) equations. These methods are well-suited for problems involving heat transfer or drag reduction, offering high accuracy and stability in their solutions. However, for FSI problems such as sloshing and swimming, grid-based methods can suffer from numerical errors due to mesh distortion when dealing with large structural movements or deformations. In contrast, mesh-free methods like smoothed particle hydrodynamics (SPH) \citep{lucy1977numerical}, moving particle semi-implicit (MPS) \citep{koshizuka1996moving}, and discrete element method (DEM) \citep{mishra1992discrete} have demonstrated significant effectiveness in capturing wave impact, breaking, and significantly varying topology. 

At present, there is no mature platform that uses mesh-free methods as the CFD environment when optimizing FSI problems with DRL. Therefore, this paper proposes an open-source Python-based platform, DRLinSPH, which utilizes the open-source library SPHinXsys \citep{zhang2021sphinxsys} and the DRL framework Tianshou \citep{weng2022tianshou} to build a parallel training platform for addressing FSI-related problems. SPHinXsys is a multi-resolution and multi-physics library based on the SPH method, which usually discretizes a continuous medium into Lagrangian particles and uses kernel functions (typically Gaussian-like functions) to approximate the mechanical interactions between them \citep{monaghan1992smoothed}. In solving incompressible fluids, the Weakly Compressible SPH (WCSPH) method is employed, using Riemann solvers to discretize the continuity and momentum equations in the NS equations \citep{zhang2017weakly}. The dual-criteria time-step method is also selected to calculate the advection and acoustic time steps, respectively \citep{zhang2020dual}. Compared with the traditional SPH methods, the calculation efficiency is improved while ensuring the calculation accuracy. Additionally, SPHinXsys is coupled with the multi-body dynamics library Simbody for the computation of rigid body kinematics and related problems \citep{zhang2021efficient}. To date, SPHinXsys has already been successfully applied to various FSI problems, including tank sloshing with baffles \citep{ren2023numerical}, wave interactions with an oscillating wave surge converter (OWSC) \citep{zhang2021efficient}, and the passive flapping of a flexible fish-like body \citep{zhang2021multi}. Another platform involved in this study is Tianshou, a DRL platform based on PyTorch and OpenAI Gym \citep{brockman2016openai}, supporting mainstream algorithms such as DQN, PPO, TD3, and SAC. It also supports vectorized environments for parallel computation and offers extensive extensibility \citep{weng2022tianshou}. Their training results based on Atari and MuJoCo significantly outperform those from platforms like OpenAI Baselines and Spinning Up.

The following sections of this paper include: Section 2 covers the fundamental theories of the SPH method and DRL, as well as the framework of the coupled platform DRLinSPH. Section 3 presents the study of four cases, including \emph{Case 1}---a comparative validation of sloshing suppression with rigid baffles by \cite{xie2021sloshing}, \emph{Case 2}---sloshing suppression using an elastic baffle, \emph{Case 3}---optimization of wave energy capture by an OWSC and \emph{Case 4}---Training of muscle-driven fish swimming in the vortices.

\section{Methodology}
\subsection{SPH methodology}
\subsubsection{Fluid model}
For incompressible viscous flow, the Lagrangian forms of the continuity equation and the momentum conservation equation are as follows
\begin{equation}
\left\{
\begin{aligned}
\label{eq:eq1}
\frac{d\rho}{dt} &= -\rho\nabla \cdot \mathbf{v} \\
\frac{d\mathbf{v}}{dt} &= -\frac{1}{\rho}\nabla p + \nu\nabla^2\mathbf{v} + \mathbf{g} + \mathbf{a}^{e} + \mathbf{a}^{sf},
\end{aligned}
\right.
\end{equation}
where $\rho$ is the density of the fluid, $\mathbf{v}$ the velocity, $p$ the pressure, $\nu$ the kinematic viscosity, $\mathbf{g}$ the gravity acceleration, $\mathbf{a}^{e}$ the external acceleration and $\mathbf{a}^{sf}$ is the acceleration acting on fluid from structure. An artificial equation of state (EoS) is used to close Equation (\ref{eq:eq1}) \citep{monaghan1994simulating}
\begin{equation}
\label{eq:eq2}
p = c^2(\rho - \rho^{0}).
\end{equation}
Here, c is the sound speed, $\rho^{0}$ the reference density, $c = 10v_{max}$ to make sure that the density varies around 1\% \citep{morris1997modeling}. $v_{max}$ is the maximum anticipated particle velocity in the flow. For \emph{Case 1--3}, $v_{\text{max}} = 2 \sqrt{gh}$, where $g = |\mathbf{g}|$, $h$ is the water depth. For \emph{Case 4}, $v_{\text{max}} = 10 v_{\text{in}}$, where $v_{\text{in}}$ represents the inlet velocity.

In SPH, the kernel approximation of the gradient field of $f(\mathbf{r})$ can be written as
\begin{equation}
\label{eq:eq3}
\nabla f(\mathbf{r}_{i}) \approx \int_{\Omega} \nabla f(\mathbf{r}_{j}) W(\mathbf{r}_{i} - \mathbf{r}_{j}, h) d\mathbf{r}_{j} = -\int_{\Omega} f(\mathbf{r}_{j}) \nabla W(\mathbf{r}_{i} - \mathbf{r}_{j}, h) d\mathbf{r}_{j},
\end{equation}
where $W(\mathbf{r}_{i} - \mathbf{r}_{j}, h)$ is the kernel function and $h$ is the smoothing length. Combining particle approximation and Taylor-expand, we can get
\begin{equation}
\label{eq:eq4}
\nabla f(\mathbf{r}_{i}) \approx - \sum_{j}[f(\mathbf{r}_{i}) + (\mathbf{r}_{j} - \mathbf{r}_{i})\cdot \nabla f(\mathbf{r}_{i})] \nabla W(\mathbf{r}_{i} - \mathbf{r}_{j}, h) V_j,
\end{equation}
where $V_{j}$ is the particle volume of particle $j$, and the 1st-order consistency is achieved if
\begin{equation}
\label{eq:eq5}
\mathbb{A}_i = -\sum_{j}\mathbf{r}_{ij}\nabla W_{ij} V_{j} \approx \mathbb{I}.
\end{equation}
Here $\mathbf{r}_{ij} = \mathbf{r}_{i} - \mathbf{r}_{j}$, $\nabla W_{ij} =\mathbf{e}_{ij} ( \partial W({r}_{ij}, h) / \partial r_{ij})$, $\mathbf{e}_{ij} = \mathbf{r}_{ij} / r_{ij}$, and $\mathbb{I}$ represents the identity matrix. However, for particles close to the boundary or irregular distributions, the 1st-order consistency will not be satisfied. In this paper, in order to maintain numerical stability, the weighted kernel gradient correction (WKGC) \citep{ren2023efficient} is used to model free-surface flow with a WKGC matrix
\begin{equation}
\label{eq:eq6}
\Bar{\mathbb{B}_i} = \omega_1 \mathbb{B}_i + \omega_2 \mathbb{I},
\end{equation}
where $\omega_1 = |\mathbb{A}_i| / (|\mathbb{A}_i| + \epsilon)$, $\omega_2 = \epsilon / (|\mathbb{A}_i| + \epsilon)$, $\mathbb{B}_i = (\mathbb{A}_i)^{-1}$, $\epsilon = \max(\alpha - |\mathbb{A}_i|, 0)$ and $\alpha = 0.5$. For a regular particle distribution, the weighted correction matrix $\Bar{\mathbb{B}_i}$ approaches $\mathbb{B}_i$, while for highly irregular particle distributions, it tends toward $\mathbb{I}$. This strategy ensures 1st-order consistency and helps reduce numerical dissipation.

Then, the discrete of Equation (\ref{eq:eq1}) can be written as \citep{zhang2017weakly}
\begin{equation}
\left\{
\begin{aligned}
\label{eq:eq7}
\frac{d\rho_{i}}{dt} &= 2\rho_{i}\sum_{j}(\mathbf{v}_{i} - \mathbf{v}^{*})\nabla W_{ij}{V_{j}} \\
\frac{d\mathbf{v}_{i}}{dt} &= -\frac{2}{\rho_{i}}\sum_{j}P^*\nabla W_{ij}V_{j} + \frac{2}{\rho_{i}}\sum_{j}\mu\frac{\mathbf{v}_{ij}}{r_{ij}}\frac{\partial W_{ij}}{\partial r_{ij}}V_{j} + \mathbf{g}_{i} + \mathbf{a}^{e}_{i} + \mathbf{a}^{sf}_{i}.
\end{aligned}
\right.
\end{equation}
Here, $\rho_{i}$ the density of particle $i$, $\mu$ the dynamic viscosity. $\mathbf{v}_{ij} = \mathbf{v}_{i} - \mathbf{v}_{j}$ the particle relative velocity, $\mathbf{v}^*$ and $P^*$ are the solutions of the Riemann problems with the piece-wise constant assumption \citep{toro2013riemann}. Typically, the one-dimensional (1D) Riemann problem in SPH can be described as
\begin{equation}
\left\{
\begin{aligned}
\label{eq:eq8}
(\rho_{L}, U_{L}, P_{L}, c_{L}) &= (\rho_{i}, -\mathbf{v}_{i} \cdot \mathbf{e}_{ij}, p_{i}, c_{i}) \\
(\rho_{R}, U_{R}, P_{R}, c_{R}) &= (\rho_{j}, -\mathbf{v}_{j} \cdot \mathbf{e}_{ij}, p_{j}, c_{j}),
\end{aligned}
\right.
\end{equation}
where subscript $L$ and $R$ mean the left and right initial states. In SPHinXsys, with a linearised Riemann solver \citep{zhang2017weakly}, the solutions can be calculated as
\begin{equation}
\left\{
\begin{aligned}
\label{eq:eq9}
\mathbf{v}^* &= \frac{\mathbf{v}_{i} + \mathbf{v}_{j}}{2} + (U^* - \frac{U_{L} + U_{R}}{2})\mathbf{e}_{ij} \\
U^{*} &=  \frac{\rho_{L}c_{L}U_{L} + \rho_{R}c_{R}U_{R} + P_{L} - P_{R}}{\rho_{L}c_{L} + \rho_{R}c_{R}} \\
P^{*} &=  \frac{\rho_{L}c_{L}P_{R}\Bar{\mathbb{B}_j} + \rho_{R}c_{R}P_{L}\Bar{\mathbb{B}_i} + \rho_{L}c_{L}\rho_{R}c_{R}\beta(U_{L} - U_{R})}{\rho_{L}c_{L} + \rho_{R}c_{R}}.
\end{aligned}
\right.
\end{equation}
Here, $\beta = \min(3\max(U_{L} - U_{R}, 0) / \overline{c}, 1)$ is the low dissipation limiter, and $\overline{c} = (\rho_{L}c_{L} + \rho_{R}c_{R}) / (\rho_{L} + \rho_{R})$.

The time integration in SPHinXsys uses a dual time-stepping approach to enhance computational efficiency. The updates of particle configuration, kernel weights and gradients, and transport-velocity formulation are governed by the advection criterion. The pressure and density relaxation and the time integration of particle density, position, and velocity are performed using a smaller time step governed by the acoustic criterion. Following \cite{zhang2020dual}, the time step size with the advection criterion $\Delta t_{ad}$ and the acoustic criterion $\Delta t_{ac}$ are
\begin{equation}
\left\{
\begin{aligned}
\label{eq:eq10}
\Delta t_{ad} &= CFL_{ad}\min(\frac{h}{v_{max}}, \frac{h^2}{\nu}) \\
\Delta t_{ac} &= CFL_{ac}(\frac{h}{c + v_{max}}),
\end{aligned}
\right.
\end{equation}
where $CFL_{ad} = 0.25$ and $CFL_{ac} = 0.6$. Besides, the particle density will be reinitialized at each advection step with
\begin{equation}
\left\{
\begin{aligned}
\label{eq:eq11}
\rho_i^{f} &= \max(\rho^*, \rho^0\frac{\sum W_{ij}}{\sum W_{ij}^0}) \\ 
\rho_i^{n} &= \rho^0\frac{\sum W_{ij}}{\sum W_{ij}^0},  
\end{aligned}
\right.
\end{equation}
where $\rho_i^{f}$ is free-surface particles, $\rho_i^{n}$ inner particles, $\rho^{0}$ the initial reference value and $\rho^*$ is the density before re-initialization.

\subsubsection{Solid model}
The mass and momentum conservation equations of the elastic structure in total Lagrangian formulations \citep{zhang2023efficient} can be established as
\begin{equation}
\left\{
\begin{aligned}
\label{eq:eq12}
\rho^s &= \rho^s_0\frac{1}{J}\\
\frac{d\mathbf{v}^s}{dt} &= \frac{1}{\rho^s} \nabla_0 \cdot \mathbb{P}^{T} + \mathbf{g} + \mathbf{a}^e + \mathbf{a}^{fs},
\end{aligned}
\right.
\end{equation}
where $\rho^s$ is the structure density, $\rho^s_0$ the initial structure density, $\mathbf{a}^{fs}$ the  acceleration acting on structure from fluid. $J = \det(\mathbb{F})$, $\mathbb{F} = \nabla_0 \mathbf{u} + \mathbb{I}$ is the deformation gradient tensor, and $\mathbf{u} = \mathbf{r}^s - \mathbf{r}^s_0$ is the displacement. $\mathbb{P}$ represents the first Piola–Kirchhoff stress tensor, $\mathbb{P} = \mathbb{F}\mathbb{S}$, $\mathbb{S}$ the second Piola–Kirchhoff stress tensor. In this paper, the structure is simplified as the linear elastic and isotropic material with Saint Venant–Kirchhoff Model, and the strain energy density function $W^s$ can be given by
\begin{equation}
\label{eq:eq13}
W^s(\mathbb{E}) = \frac{\lambda^s}{2}(\text{tr}(\mathbb{E}))^2 + \mu^s\text{tr}(\mathbb{E}^2).
\end{equation}
Here, $\mathbb{E} = (\mathbb{F}^{T}\mathbb{F} - \mathbb{I}) / 2$ is the Green–Lagrange strain tensor. $\lambda^s$ and $\mu^s$ are the Lamé parameters with
\begin{equation}
\left\{
\begin{aligned}
\label{eq:eq14}
\lambda^s &= \frac{E \nu^s}{(1 + \nu^s)(1-2\nu^s)}\\
\mu^s &= \frac{E}{2(1 + \nu^s)},
\end{aligned}
\right.
\end{equation}
where $\nu^s$ is the Poisson’s ratio, $E$ the Young’s modulus, and $\mathbb{S}$ can be obtained with
\begin{equation}
\label{eq:eq15}
\mathbb{S} = \frac{\partial W^s}{\partial \mathbb{E}} = \lambda^s\text{tr}(\mathbb{E}) + 2\mu^s\mathbb{E}
\end{equation}

The kernel gradient correction matrix $\mathbb{B}_{i}^0$ for structure based on Equation \ref{eq:eq6} ($\epsilon = 0$) \citep{vignjevic2006sph} is calculated from the initial reference configuration with
\begin{equation}
\label{eq:eq16}
\mathbb{B}_{i}^0 = (- \sum_{j} (\mathbf{r}_{ij}^s)_0 \nabla_0 W_{ij}^s V_j^s)^{-1},
\end{equation}
where $\nabla_0 W_{ij}^s =(\mathbf{e}_{ij}^{s})_0 ( \partial W(({r}_{ij}^s)_0, h^s) / \partial ({r}_{ij}^s)_0)$, $h^s$ is the smooth length used for the structure. 

Then, the discrete of Equation (\ref{eq:eq12}) can be written as \citep{zhang2022smoothed}
\begin{equation}
\left\{
\begin{aligned}
\label{eq:eq17}
\rho^s_i &= \rho^s_0\frac{1}{\det(\mathbb{F}_i)}\\
\frac{d\mathbf{v}^s_i}{dt} &= \frac{1}{\rho^s_i} \sum_j (\mathbb{P}_i \mathbb{B}_{i}^0 + \mathbb{P}_j \mathbb{B}_{j}^0) \nabla_0 W_{ij}^s V_j^s + \mathbf{g}_i + \mathbf{a}^e_i + \mathbf{a}^{fs}_i,
\end{aligned}
\right.
\end{equation}
where $\mathbb{F}_i$ is discretized as
\begin{equation}
\label{eq:eq18}
\mathbb{F}_i = (- \sum_{j} \mathbf{r}_{ij}^s \nabla_0 W_{ij}^s V_j^s)\mathbb{B}_{i}^0 + \mathbb{I}.
\end{equation}

In the simulation of muscle movements by applying active strain $\mathbb{E}_a$ to the structure \citep{nardinocchi2007active, curatolo2016modeling}, the deformation gradient tensor $\mathbb{F}_t $ is typically modified as $\mathbb{F}_t = \mathbb{F} \mathbb{F}_a$. $\mathbb{F}_a$ is a time-varying tensor field and $\mathbb{E}_a = (\mathbb{F}_a^{T}\mathbb{F}_a - \mathbb{I}) / 2$. The first Piola–Kirchhoff stress tensor $\mathbb{P}_t$ is changed to $\mathbb{P}_t = \mathbb{P} \mathbb{F}_a^*$, $\mathbb{F}_a^* = \det(\mathbb{F}_a)(\mathbb{F}_a^{-1})^T$.

Besides, the time step for structure \citep{zhang2021multi} is
\begin{equation}
\label{eq:eq19}
\Delta t^{s} = 0.6\min(\frac{h^s}{c^s + v_{max}}, \sqrt{\frac{h^s}{(dv/dt)_{max}}}).
\end{equation}
Here, $c^s = \sqrt{K/\rho^s}$ is the structure sound speed, $K = \lambda^s +2 \mu^s/3$ the bulk modulus.

\subsubsection{Fluid-structure coupling}
In SPHinXsys, the structure is treated as the moving wall boundary for fluid. The one-side Riemann problem is constructed along the structure for solving the continuity and momentum equations \citep{zhang2022smoothed}. The total force from structure $\mathbf{a}^{sf}$ can be written as
\begin{equation}
\label{eq:eq20}
\mathbf{a}^{sf} = -\frac{2}{\rho_i}\sum_a P^* \nabla W_{ia} V_a + \frac{2}{\rho_i}\sum_a \mu \frac{\mathbf{v}_{ia}}{r_{ia}}\frac{\partial W_{ia}}{\partial r_{ia}}V_{a},
\end{equation}
where $a$ represents the structure particle. The smooth length $h$ in $W_{ia}$ is from fluid with the assumption $h>h^s$. Besides the first item on the right is the pressure force, and the second is the viscous force. $P^*$ is also calculated from Equation (\ref{eq:eq9}). Considering that $\Delta t^{s} < \Delta t_{ac}$ due to $c^s > c$, there will be a force mismatch problem in FSI. \cite{zhang2021multi} used $\Bar{\mathbf{v}}_a$ which is the averaged velocity of the structure particle within a fluid acoustic time step for the calculation, and $\mathbf{v}_{ia} = \mathbf{v}_i - \Bar{\mathbf{v}}_a$. The left and right sides of the one-side Riemann problem can be present as
\begin{equation}
\left\{
\begin{aligned}
\label{eq:eq21}
(\rho_{L}, U_{L}, P_{L}) &= (\rho_{i}, -\mathbf{v}_{i} \cdot \mathbf{n}_{a}, p_{i}) \\
(\rho_{R}, U_{R}, P_{R}) &= (\rho_{a}, -(2\mathbf{v}_{i}-\Bar{\mathbf{v}}_a) \cdot \mathbf{n}_{a}, p_{a}).
\end{aligned}
\right.
\end{equation}
Here, $\mathbf{n}_{a}$ is the normal vector from the structure to the fluid, and $p_{a}$ can be described as
\begin{equation}
\label{eq:eq22}
p_{a} = p_{i} + \rho_i \max(0, (\mathbf{g} - d\Bar{\mathbf{v}}_a / dt)\cdot \mathbf{n}_{a})(\mathbf{r}_{ia}\cdot\mathbf{n}_{a}).
\end{equation}

\begin{figure}[ht]
\centering
\includegraphics[width=\textwidth]{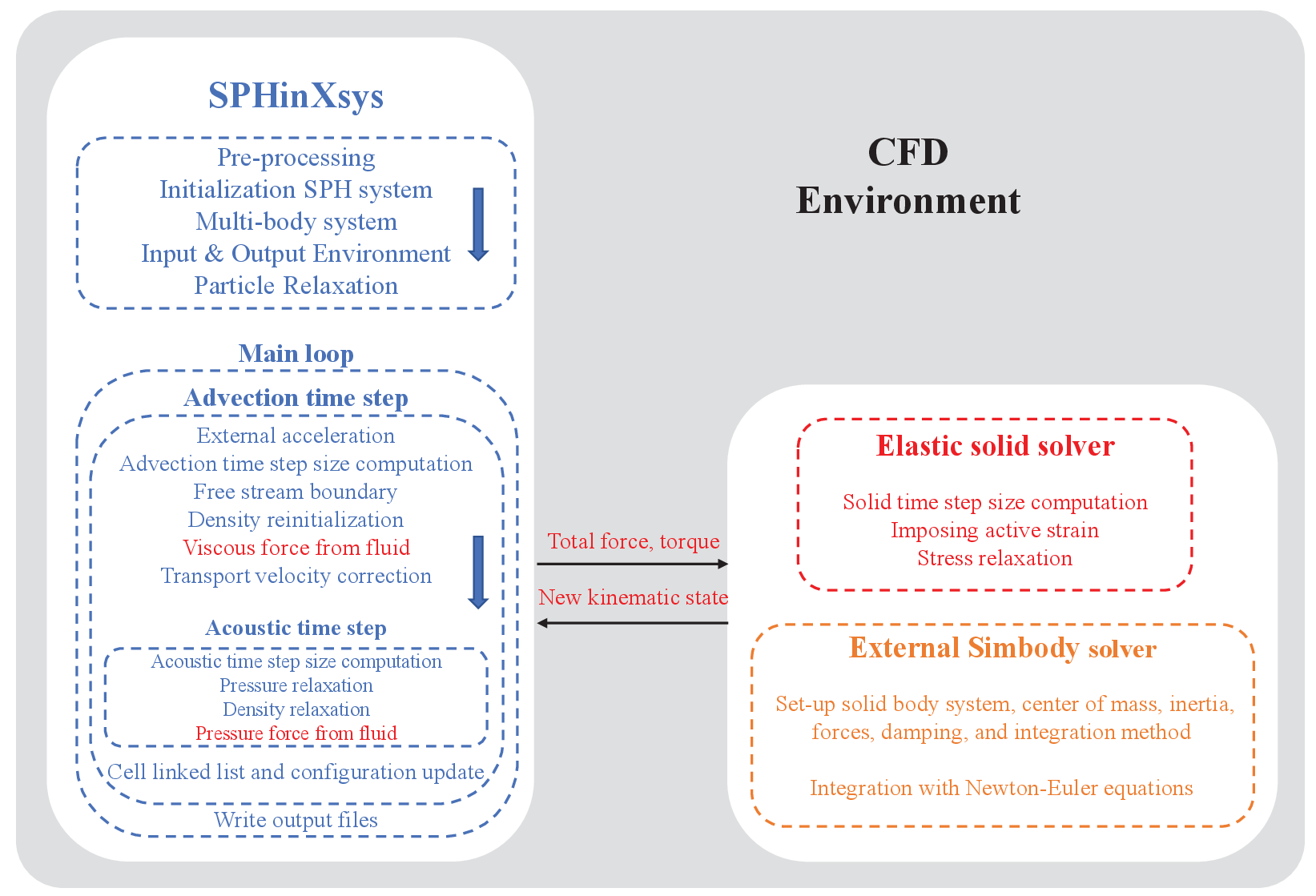}
\caption{The flowchart for solving FSI problems under SPHinXsys. The solution for the elastic structure is handled by SPHinXsys, while rigid body kinematics are addressed using Simbody.}\label{fig1}
\end{figure}

Additionally, for problems related to rigid bodies, such as the rotational motion of the flap hinged at the bottom, the forces and torque computed in SPHinXsys are transferred to Simbody for solving the kinematics \citep{zhang2021efficient} as shown in Figure \ref{fig1}.

\subsection{DRL algorithms}
DRL algorithms can be generally classified into on-policy and off-policy methods. On-policy algorithms, such as PPO \citep{schulman2017proximal}, update their policy after each episode and use the newly updated policy to gather data in the next episode. In contrast, off-policy algorithms use different policies for updating and data collection, with SAC \citep{haarnoja2018soft} being a typical example. Furthermore, most of the current DRL algorithms adopt an actor-critic architecture, where the actor represents the policy network, and the critic typically evaluates the performance of the policy based on the action-value function or state-value function \citep{sutton2018reinforcement}.

The action value function $Q_{\pi_\theta}(s_{n}, a_{n})$ is defined as
\begin{equation}
\label{eq:eq23}
Q_{\pi_\theta}(s_{n}, a_{n}) = \mathbf{E}_{\pi_\theta}[\sum_{t=n}^{\infty} (\gamma_{t} r_{t} | s_{t}, a_{t})].
\end{equation}
Here, $\mathbf{E}$ denotes the conditional expectation given the observed state $s_n$ and action $a_n$. $\pi_{\theta}$ represents the policy network with parameter $\theta$, which takes the state $s_t$ as input and the output is action $a_t$. The term $\sum_{t=n}^{\infty} \gamma_{t} r_{t}$ is commonly referred to as the return $U_{n}$ where $\gamma_{t} \in [0, 1]$ is the discount factor that weights future rewards. The $Q$ increases with more favorable state-action pairs $(s_n, a_n)$ and an improved policy $\pi_{\theta}$.

The state value function $V_{\pi_\theta}(s_{n})$ can be written with
\begin{equation}
\label{eq:eq24}
V_{\pi_\theta}(s_{n}) = \mathbf{E}_{a_{n} \sim \pi_\theta}[\sum_{t=n}^{\infty} (\gamma_{t} r_{t} | s_{t})].
\end{equation}
This function differs from $Q_{\pi_\theta}(s_{n}, a_{n})$ in that it takes the expectation over the actions at time step $n$, making $V$ dependent only on $s_n$ and $\pi_{\theta}$. As the current state $s_n$ and $\pi_{\theta}$ improve, the value of $V$ increases.

\subsubsection{Proximal Policy Optimization Algorithm}
The core of the PPO algorithm is constructing an objective function $J(\theta)$. The objective function typically represents the return \(U_{n}\) as a function of the parameter \(\theta\) of the current policy network. The optimal policy and corresponding return can be achieved by iteratively updating \(\theta\) to maximize the objective function. Based on the policy gradient theorem (PGT), the objective function $J(\theta)$ can be written as \citep{silver2014deterministic, schulman2017proximal}
\begin{equation}
\label{eq:eq25}
J(\theta) = \mathbf{E}_{s_{n} \sim \mathcal{D}}[\mathbf{E}_{a_{n} \sim \pi_{\theta_m}}[\frac{\pi_{\theta}(a_{n}|s_{n})}{\pi_{\theta_{m}}(a_{n}|s_{n})}\cdot A^{\pi_{\theta_m}}(s_{n}, a_{n})]].
\end{equation}
Here, $\mathcal{D}$ is the replay buffer, and $\theta_{m}$ denotes the parameters of the policy network from the previous iteration. $A^{\pi_{\theta_m}}(s_{n}, a_{n})$ is the advantage function \citep{schulman2015high}, defined as
\begin{equation}
\label{eq:eq26}
A^{\pi_{\theta_m}}(s_{n}, a_{n}) = r_{n} + \gamma V^{\pi_{\theta_m}}_{\phi}(s_{n + 1}) - V^{\pi_{\theta_m}}_{\phi}(s_{n}),
\end{equation}
where $V^{\pi_{\theta_m}}_{\phi}(s_n)$ is the value of state $s_n$ modeled using a critic network parameterized by $\phi$, which is represented by DNNs. The superscript $\pi_{\theta_m}$ indicates that the data is collected using the policy from the $m$-th iteration, and $\gamma$ is the discount factor. While the advantage function does not change the overall expectation, it enhances the performance of the policy \cite{schulman2015high}. Besides, the clipped surrogate objective is imposed in PPO and is designed to prevent huge updates to the policy \citep{schulman2017proximal}. 

The objective function $J(\theta)$ is finally reformulated as follows
\begin{align}
\label{eq:eq27}
J(\theta) &= \mathbf{E}_{s_{n} \sim \mathcal{D}} \left[ \mathbf{E}_{a_{n} \sim \pi_{\theta_m}} \left[ \min \left( r_{\theta}(s_{n}, a_{n}), \right. \right. \right. \notag \\
& \left. \left. \left. \text{clip}(r_{\theta}(s_{n}, a_{n}), 1 - \sigma, 1 + \sigma) \right) \cdot A^{\pi_{\theta_m}}(s_{n}, a_{n}) \right] \right],
\end{align}
where $r_{\theta}(s_{n}, a_{n}) = \pi_{\theta}(a_{n}|s_{n}) / \pi_{\theta_{ m}}(a_{n}|s_{n})$, $\sigma = 0.2$. 

The loss function for the critic network $V_{\phi}$ is based on the TD method, defined as
\begin{equation}
\label{eq:eq28}
L(\phi) = \mathbf{E}_{s_{n} \sim \mathcal{D}} \left[ \left( V_{\phi}(s_{n}) - (r_n + \gamma V_{\phi}(s_{n+1})) \right)^2 \right].
\end{equation}

The policy network $\pi_{\theta}$ is updated by maximizing $J(\theta)$ with stochastic gradient ascent, while the critic network is $V_{\phi}$ using gradient descent to minimize the mean squared errors, both implemented with the Adam optimizer \cite{kingma2014adam}
\begin{equation}
\label{eq:eq29}
\left\{
\begin{aligned}
\theta_{m + 1} &\leftarrow \theta_{m} + \alpha \nabla_\theta J(\theta) \\
\phi_{n+1} &\leftarrow \phi_{n} - \beta \nabla_\phi L(\phi),
\end{aligned}
\right.
\end{equation}
where $\alpha$ and $\beta$ are the learning rates.

\subsubsection{Soft Actor Critics Algorithm}
In the PPO algorithm, the randomness in exploration primarily arises from sampling actions based on the probability density function output by the policy network. In contrast, algorithms such as TD3 introduce exploration by directly adding noise to the action outputs \citep{fujimoto2018addressing}. The SAC algorithm, on the other hand, incorporates the policy's entropy into the state-value function, promoting exploration by maximizing the entropy-regularized return \citep{haarnoja2018soft}. The state value function $V_{\pi_\theta}(s_{n})$ can be rewritten with
\begin{equation}
\label{eq:eq30}
V_{\pi_\theta}(s_{n}) = \mathbf{E}_{a_{n} \sim \pi_\theta}[\sum_{t=n}^{\infty} (\gamma_{t} r_{t} | s_{t}) + \chi\mathcal{H}(\pi_{\theta}(s_{n}))].
\end{equation}
Here, $\chi$ is the entropy coefficient.

\begin{figure}[htbp]
\centering
\includegraphics[width=\textwidth]{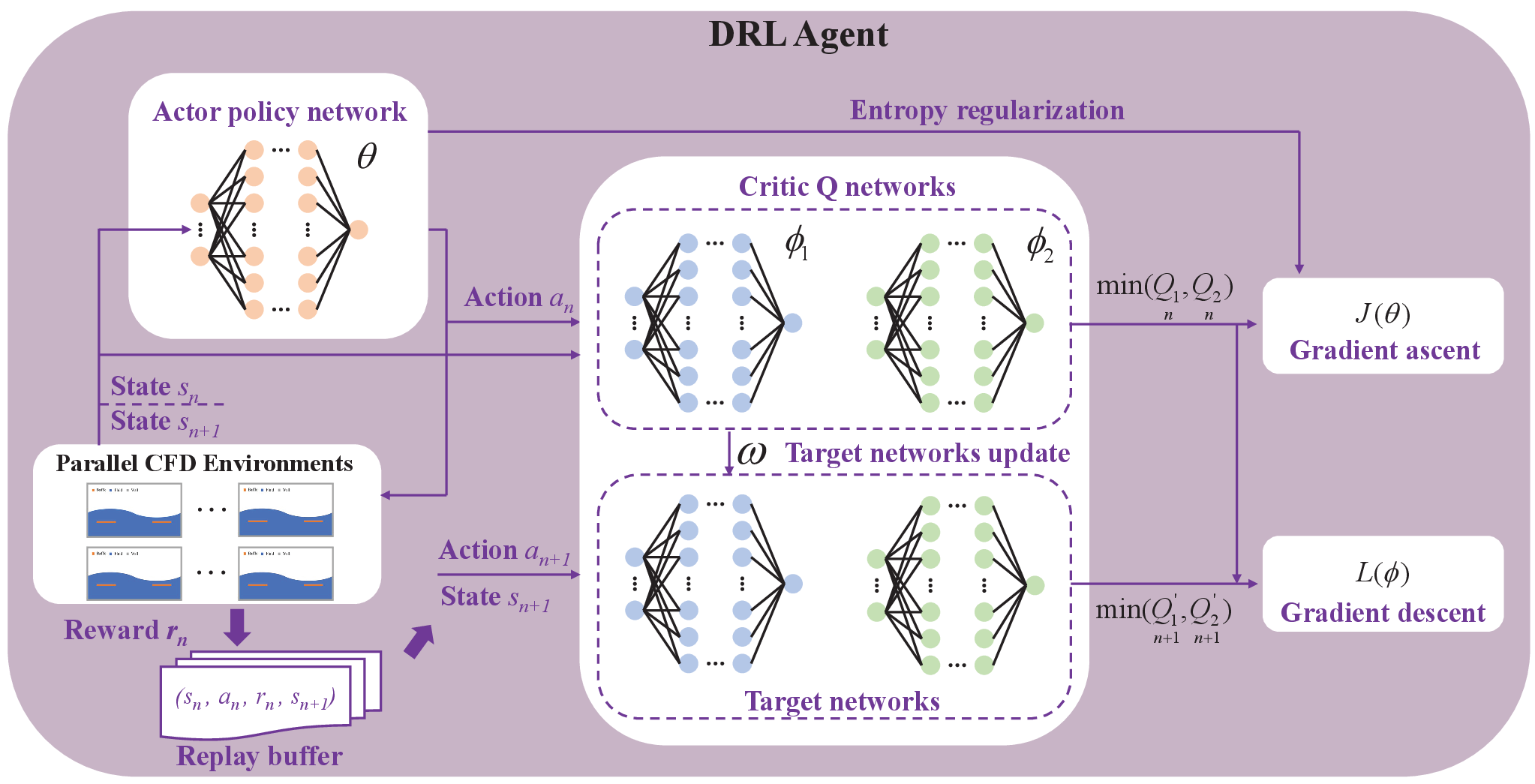}
\caption{The structure of the DRL agent based on the SAC algorithm and its interaction with the parallel CFD environment.}\label{fig2}
\end{figure}

Compared to PPO, SAC employs five DNNs: a policy network, two critic networks, and two target networks corresponding to the critic networks, as shown in Figure \ref{fig2}. Unlike PPO, where the critic network is based on the state-value function $V_{\phi}(s_{n})$, SAC utilizes the action-value function $Q_{\phi}(s_{n}, a_{n})$ for the critic networks. So, the target networks $Q_{\phi^{'}}(s_{n}, a_{n})$ are necessary to help mitigate issues such as value overestimation and training instability, providing more stable target estimates during the learning process. 

The objective function $J(\theta)$ for the policy network $\pi_{\theta_m}$ is computed as
\begin{equation}
\label{eq:eq31}
J(\theta) = \mathbf{E}_{s_n \sim \mathcal{D}} \left[\min\limits_{i=1,2} Q_{\phi_{i}}(s_n, \tilde{a}_{n}) - \chi \log \pi_\theta(\tilde{a}_{n} | s_n) \right],
\end{equation}
where $\tilde{a}_{n}$ is the sample from $\pi_{\theta}(\cdot | s_{n})$.

The loss function for the critic networks $Q_{\phi}$ is also calculated by TD methods
\begin{equation}
\label{eq:eq32}
L(\phi_i) = \mathbf{E}_{s_{n} \sim \mathcal{D}} \left[ \left( Q_{\phi_i}(s_{n}, a_{n}) - y_{n} \right)^2 \right], i = 1, 2. 
\end{equation}
Here $y_{n} = r_{n} + \gamma \left( \min_{i=1,2} Q_{\phi_i'}(s_{n+1}, \tilde{a}_{n+1}) - \chi \log \pi_{\theta}(\tilde{a}_{n+1} | s_{n+1}) \right)$, $\tilde{a}_{n+1}$ is the sample from $\pi_{\chi}(\cdot | s_{n+1})$.

The network updates in SAC are similar to those in PPO, except for the target networks, which are updated using a weighting factor $\omega \in [0, 1]$
\begin{equation}
\label{eq:eq33}
(\phi_i')_{n+1} \leftarrow \omega(\phi_i')_{n} + (1-\omega)(\phi_i)_n, i = 1, 2. 
\end{equation}

Considering that SAC offers notable advantages in terms of stability, sample efficiency, and robustness in continuous control tasks and has already seen widespread application \citep{liang2024environmental, zhang2024dual, cui2024enhancing}, this paper will primarily focus on employing the SAC algorithm.

\subsection{DRLinSPH}
\begin{figure}[htbp]
\centering
\includegraphics[width=\textwidth]{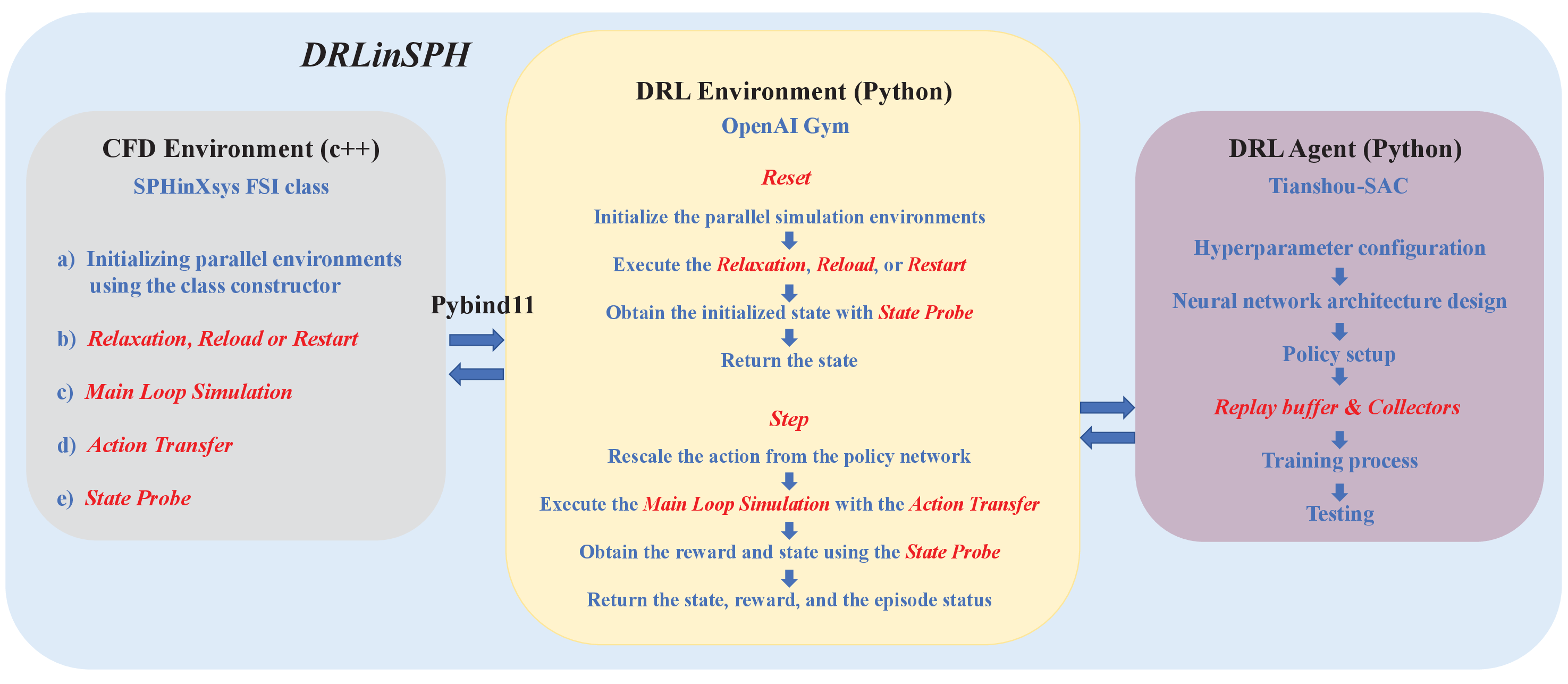}
\caption{The structure of DRLinSPH is composed of three key components: the CFD environment, the DRL environment, and the DRL agent.}\label{fig3}
\end{figure}

To effectively control and optimize FSI problems using DRL, it is critical to integrate Python-based DRL platforms, such as Tianshou, with C++-based CFD environments, like SPHinXsys through DRLinSPH. As illustrated in Figure \ref{fig3}, this paper first establishes a comprehensive environment (class) within SPHinXsys for the specific FSI problem, incorporating necessary solvers for fluid and solid dynamics. Subsequently, four essential member functions are defined within this class: The first is the \emph{Relaxation}, \emph{Reload} and \emph{Restart}, which handle particle relaxation and reloading, or restarting simulations from a specific time step. The \emph{Restart} is handy in cases where simulations require an initial period to reach numerical stability before commencing optimization. The second is the \emph{Main Loop Simulation}, responsible for the numerical computations. The third is the \emph{Action Transfer}, which facilitates the real-time transmission of actions $a_{n}$ derived from the DRL agent to the numerical solver. Finally, the \emph{State Probe} extracts necessary data for the state $s_n$, such as velocity and pressure at specific points within the flow field.

Following the setup of the SPHinXsys environment, a custom DRL environment based on OpenAI Gym \citep{brockman2016openai} is developed. This environment includes two core functions: \emph{Reset} and \emph{Step}. Since SPHinXsys is compatible with various system platforms, including Unix-like and Windows systems, Pybind11 compiles the SPHinXsys FSI class into a shared object (SO) or dynamic-link library (DLL) \citep{pybind11}. It will be directly initialized within the DRL environment's \emph{Reset} through Python's import mechanism, while the initial state $s_0$ is also retrievable via the \emph{State Probe}. The \emph{Step} executes the actions $a_n$ determined by the agent, rescaling the actions derived from DNNs before passing them into the numerical solver through the \emph{Action Transfer}, enabling real-time dynamic control. At the end of each action step, a reward is calculated based on its definition, and the next state $s_{n+1}$ is retrieved via the \emph{State Probe}, completing one interaction cycle between the environment and the agent.

The Tianshou platform is utilized for the entire DRL training process \citep{weng2022tianshou}. Taking the SAC algorithm as an example, the first step involves setting algorithm-related hyperparameters and configuring the architecture of the DNNs based on the dimensions of the state and action spaces. The algorithm is then set up, and data collection is carried out through Collectors, which call the \emph{Reset} and \emph{Step} in the DRL environment. A replay buffer stores data $(s_n, a_n, r_n, s_{n+1})$ for both parallel and single-environment setups. Finally, the training process is executed, with testing performed at the end of each epoch to evaluate the performance of the learned policy.

\section{Case Studies}
\subsection{Case 1: Sloshing suppression with rigid baffles}
\begin{figure}[htbp]
\centering
\includegraphics[width=0.7\textwidth]{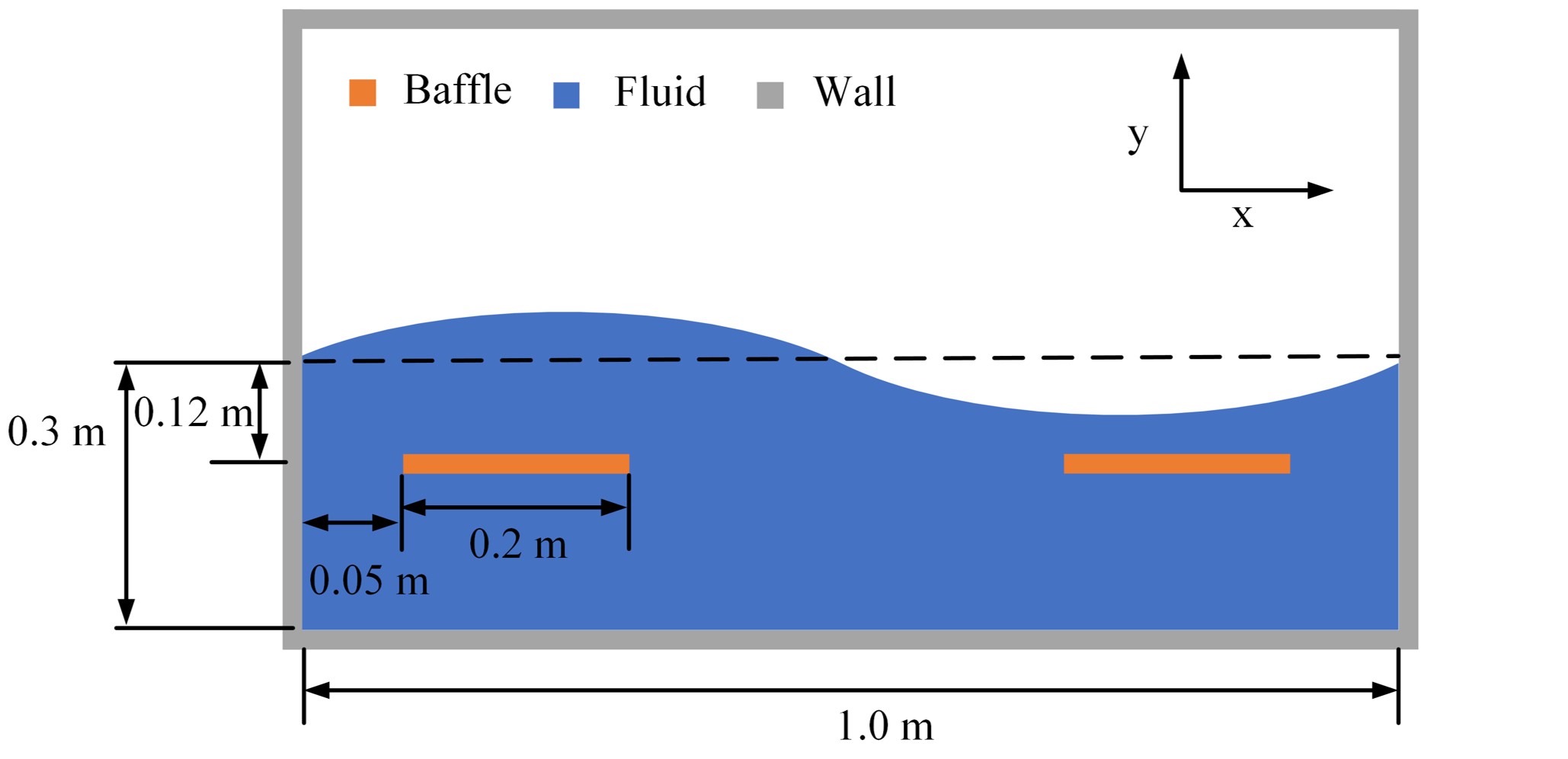}
\caption{The configuration for the case of liquid sloshing in a tank involves active-controlled baffles moving vertically.}\label{fig4}
\end{figure}

The first case in this study is based on the work by \cite{xie2021sloshing}, which investigates sloshing suppression in a 2D rectangular tank with two active-controlled baffles, as shown in Figure \ref{fig4}. The tank has a length of 1.0 m, and the water depth is 0.3 m. Two baffles are symmetrically positioned along the centerline of the tank. Each baffle is placed 0.12 m below the water surface and 0.05 m from the tank walls. The movement of the tank in the x-direction can be calculated with
\begin{equation}
\label{eq:eq34}
x = X\sin{\omega_e t},
\end{equation}
where $X = 0.002 \ \text{m}$ is the amplitude, $\omega_e = 4.762 \ \text{rad/s}$ the excitation frequency of the sloshing tank, which equals the natural frequency $\omega_0$ of the corresponding tank without baffles.

In the work of \cite{xie2021sloshing}, eight probes were used to capture the state $s_n$, including the position and velocity of the baffles, as well as the wave height and wave surface velocity. Drawing from the experiments conducted by \cite{rabault2019artificial}, which demonstrated that increasing the number of probes improves the agent's performance, this paper increases the number of probes to 37. The additional observations include the free surface height and velocity at eleven uniformly distributed measurement points along the x-direction of the tank.

\cite{xie2021sloshing} conducted a comparative study using the PPO and TD3 algorithms and concluded that TD3 resulted in shorter training times and superior performance. However, it is essential to note that although TD3 successfully controlled the baffle's motion and significantly suppressed sloshing in their study, a significant issue arose due to large changes in the baffle's velocity over short intervals. This problem stemmed from two possible reasons: the agent's action $a_n$ was directly defined as the baffle's velocity $v_y$, and the TD3 algorithm tended to produce actions near the boundary values. These rapid fluctuations led to numerical instabilities and divergence in the simulations, making the strategy unsuitable for real-world engineering applications. To address this issue, we optimized the action $a_n$ by switching to the change in velocity $\Delta v_y$, with a restriction $|\Delta v_y| \leq 0.03 \ \text{m/s}$ in one action time step $t_a = 0.1 \ \text{s}$. The update of action in the simulation is obtained for each baffle with
\begin{equation}
\label{eq:eq35}
c_{i+1} = c_i + \frac{\Delta v_y}{N}.
\end{equation}
Here, $c_i$ is the value at previous numerical time step $t_i$ and $c_{i+1}$ is the new step, $t_{i+1} - t_i \approx 12 \Delta t_{ad}$, $N = 30$.

Furthermore, the reward for each action step $r_n$ has been modified as
\begin{equation}
\label{eq:eq36}
r_n = 1 - \frac{|\eta_l - \eta_r|}{H} - p_0 - p_1,
\end{equation}
where $\eta_l$ is the free surface height at left wall, $\eta_r$ the free surface height at right wall, $H = 0.02$ m. $p_0$ and $p_1$ are the penalty for the baffle's velocity $v_y$ and distance $\Delta Y$ between current position and initial position with
\begin{equation}
p_0 = 
\begin{cases} 
-1, & \text{if } |{v}_{yl}| \text{ or } |{v}_{yr}| > 0.06 \, \text{m/s}, \\
0, & \text{otherwise}
\end{cases}
\label{eq:eq37}
\end{equation}
\begin{equation}
p_1 = 
\begin{cases} 
-10, & \text{if } |\Delta Y_l| \text{ or } |\Delta Y_r| > 0.05 \, \text{m}, \\
0, & \text{otherwise}.
\end{cases}
\label{eq:eq38}
\end{equation}
Here, subscript $l$ and $r$ mean left and right baffles.

The training for each episode will begin at \(24 \ \text{s}\) with \emph{Restart} as the motion inside the tank stabilizes. Each episode will terminate after 200 action time steps or upon meeting the condition \(p_1\). Two algorithms, PPO and SAC, are used for comparison. All computations in this paper were performed on the Mac OS system equipped with two Apple M1 Max cores and 64 GB of RAM. The remaining hyperparameters of the algorithms used in this case are listed in Table \ref{table1}.
\begin{table}[htbp]
    \centering
    \caption{Basic hyperparameters of different DRL algorithms.}
    \label{table1}
    \begin{tabular}{@{}lccc@{}}
        \toprule
        Algorithm & PPO & SAC \\ \midrule
        Network structure & [512, 512] & [512, 512] \\
        Activation function & \(\tanh\) & \(\tanh\) \\ 
        Learning rate (\(\alpha\)) & 3e-4 & 1e-3 \\ 
        Steps per epoch & 2048 & 2048 \\ 
        Batch size & 256 & 256 \\ 
        Discount factor (\(\gamma\)) & 0.99 & 0.99 \\ 
        Soft update (\(\omega\)) & - & 0.005 \\ \bottomrule
    \end{tabular}
\end{table}

\subsubsection{Numerical model validation}
\begin{figure}[htbp]
\centering
\includegraphics[width=0.6\textwidth]{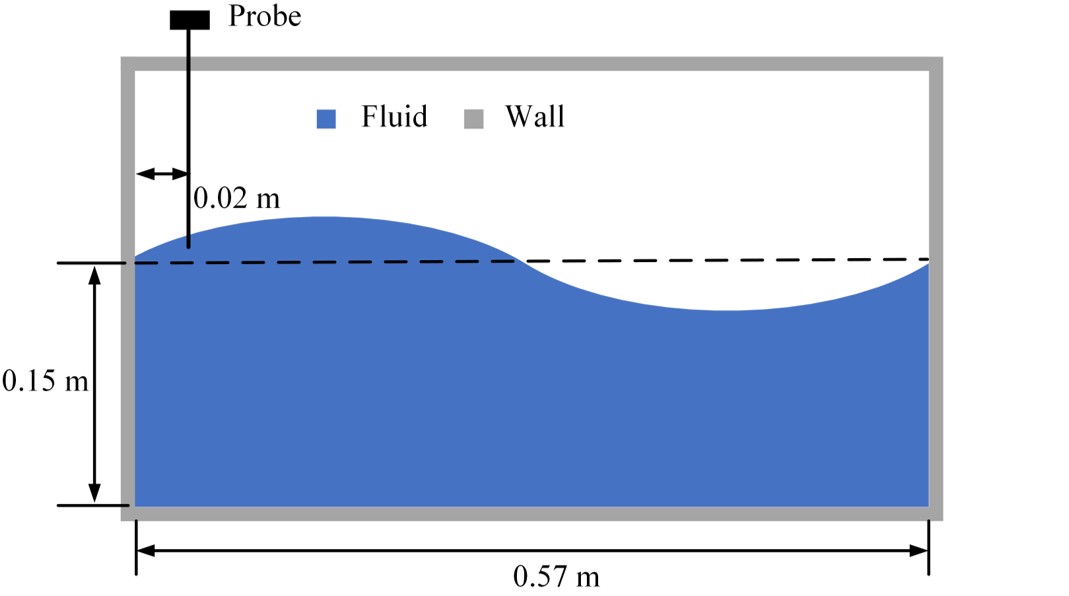}
\caption{The geometry of the tank without baffles.}\label{fig5}
\end{figure}
\begin{figure}[htbp]
\centering
\includegraphics[width=\textwidth]{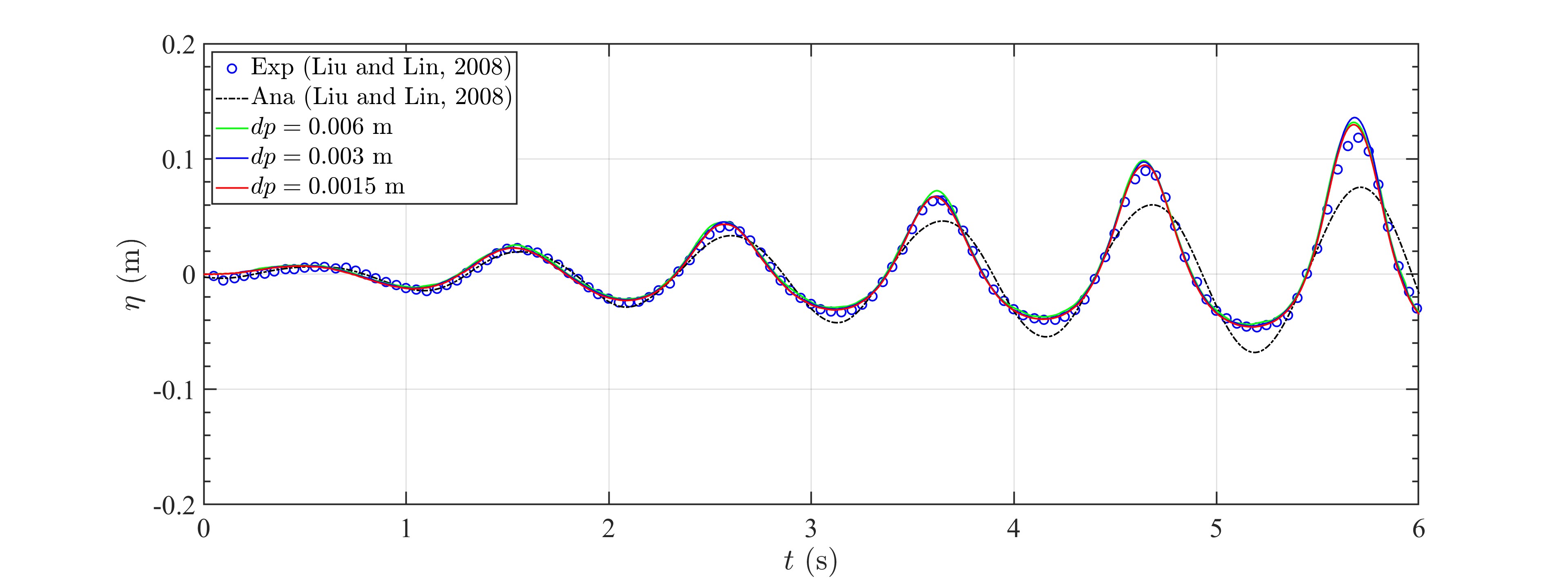}
\caption{Comparisons of free surface elevation at $x = 0.02 \ \text{m}$ with different particle resolutions and the experiment.}\label{fig6}
\end{figure}
Two benchmarks were employed to validate the accuracy of the numerical model before training. The first is the sloshing in a tank without baffles in Figure \ref{fig5} \citep{liu2008numerical}. The 2D tank has a length $L$ of 0.57 m and a water depth $h$ of 0.3 m. The excitation frequency $\omega_e$ is set to 6.0578 $\text{rad/s}$ with an amplitude of $X = 0.005 \ \text{m}$. A free surface height probe is put near the left wall at a distance of 0.02 m. The results for different fluid particle resolutions $dp$ and experimental and analytical data are presented in Figure \ref{fig6}. It can be observed that the numerical simulation accurately captures both the period and amplitude of the free surface near the wall. The results for resolutions of $dp = 0.003 \ \text{m}$ and $dp = 0.0015 \ \text{m}$ show minimal differences. Therefore, a resolution of $dp = 0.003 \ \text{m}$ is used for training. 
\begin{figure}[htbp]
\centering
\includegraphics[width=0.7\textwidth]{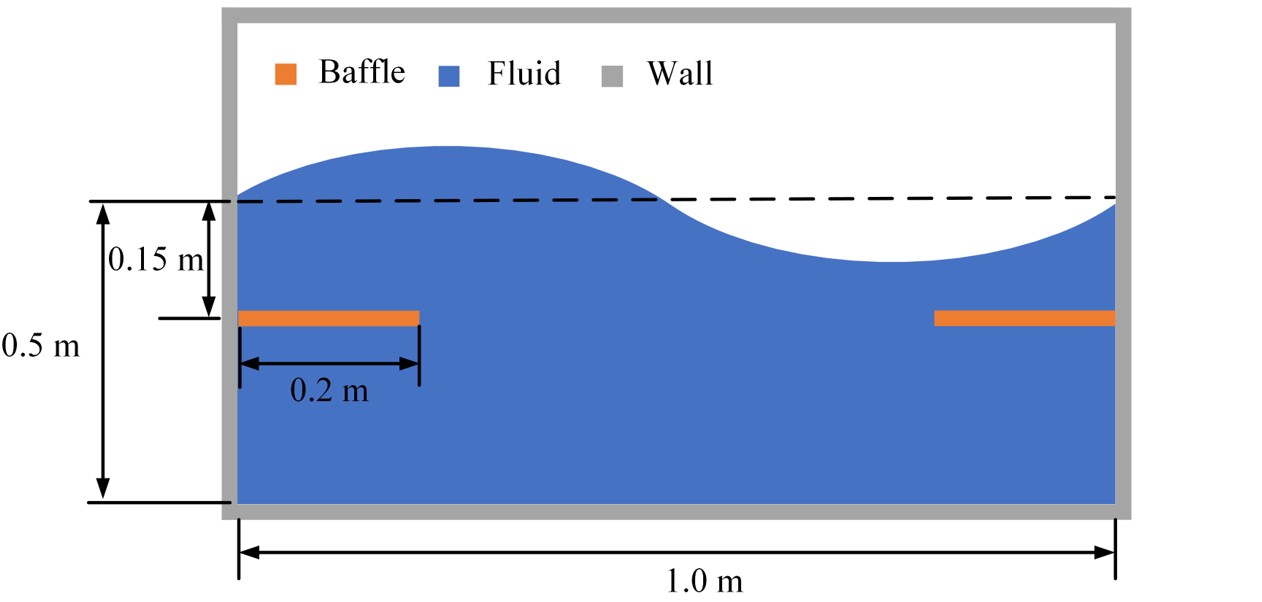}
\caption{The geometry of the tank with baffles.}\label{fig7}
\end{figure}

\begin{figure}[htbp]
\centering
\includegraphics[width=\textwidth]{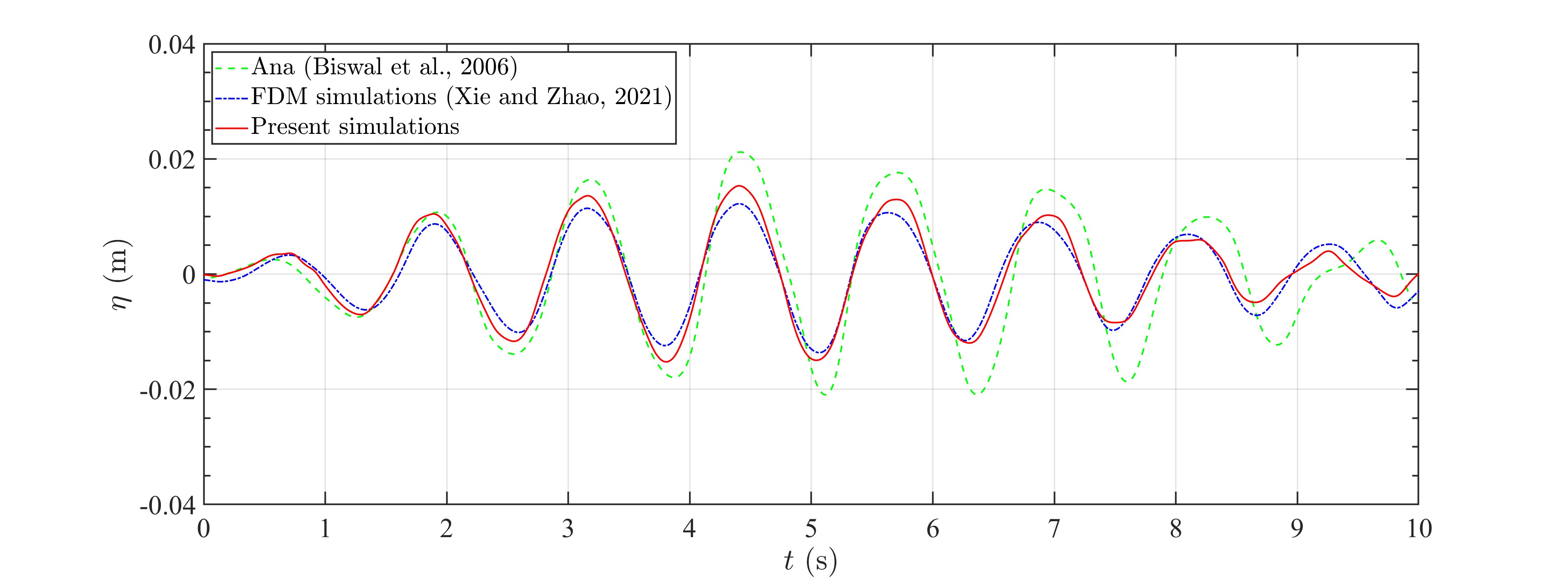}
\caption{Comparisons of the free surface elevation at the right wall between the analytical results and simulations.}\label{fig8}
\end{figure}

The second case involves sloshing in a tank equipped with two fixed, rigid baffles \citep{biswal2006non}, as shown in Figure \ref{fig7}. Each baffle has a width of 0.2 m and is positioned 0.15 m below the free surface. The tank has a length of 1.0 m and a water depth of 0.3 m. The excitation frequency $\omega_e$ is set to $5.29 \ \text{rad/s} = 0.995 \omega_0$, with an amplitude of $X = 0.002 \ \text{m}$. From Figure \ref{fig8}, we can see that the results of the numerical simulations are in close agreement with those of \cite{xie2021sloshing} in terms of amplitude and period.

\subsubsection{Results}
\begin{figure}[htbp]
\centering
\includegraphics[width=\textwidth]{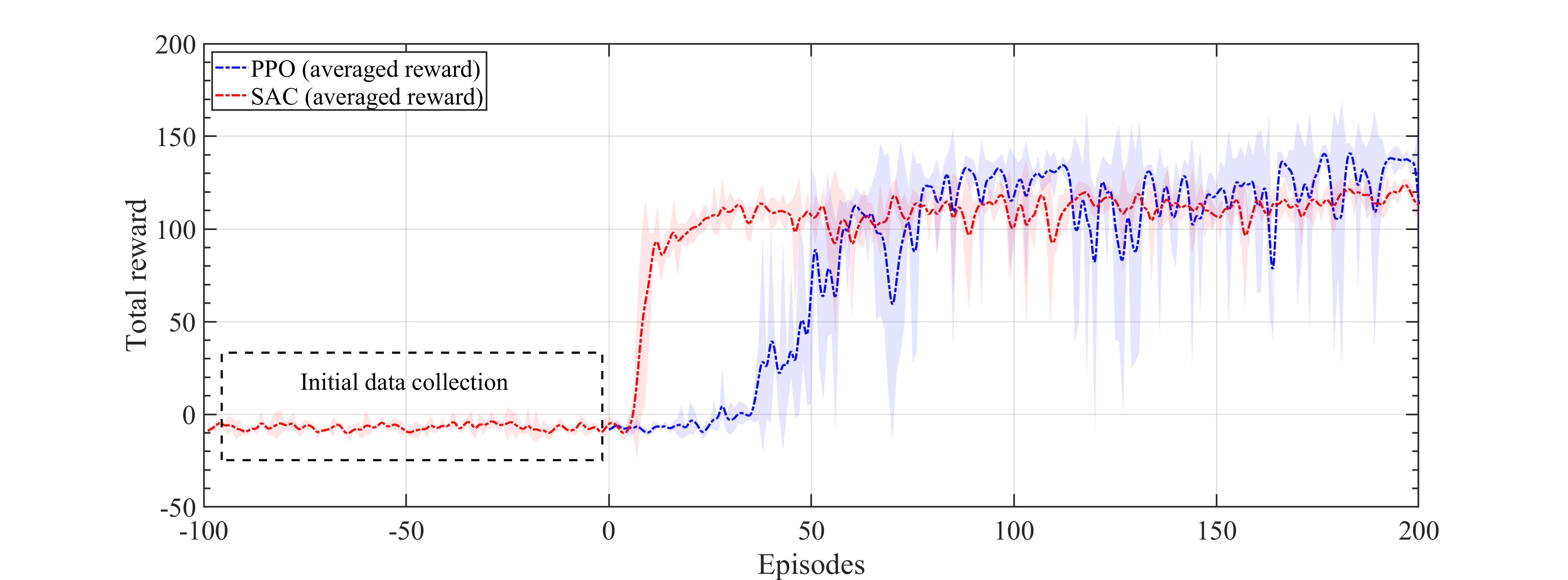}
\caption{The training curves of the PPO and SAC algorithms illustrate the relationship between episodes and the total reward per episode. The dashed line represents the average total reward across parallel environments, while the shaded area indicates the standard deviation of the total reward.}\label{fig9}
\end{figure}

The training curves for the two algorithms are presented in Figure \ref{fig9}. The agent trained using the SAC algorithm demonstrates a faster ability to explore and identify more effective strategies than those trained with the PPO algorithm. This advantage arises because SAC collects substantial data about approximately two epochs with stochastic noise before updating its policy, enabling a more thorough exploration of the state-action space. Moreover, the standard deviation of the total reward is markedly lower for the SAC algorithm, suggesting two critical points: first, the entropy-regularized nature of SAC promotes stability, and second, as an off-policy algorithm, SAC leverages the benefits of a replay buffer, which stores all the data collected by the agent during the training, allowing the algorithm to reuse past experiences for multiple updates to the policy.

\begin{figure}[htbp]
\centering
\includegraphics[width=\textwidth]{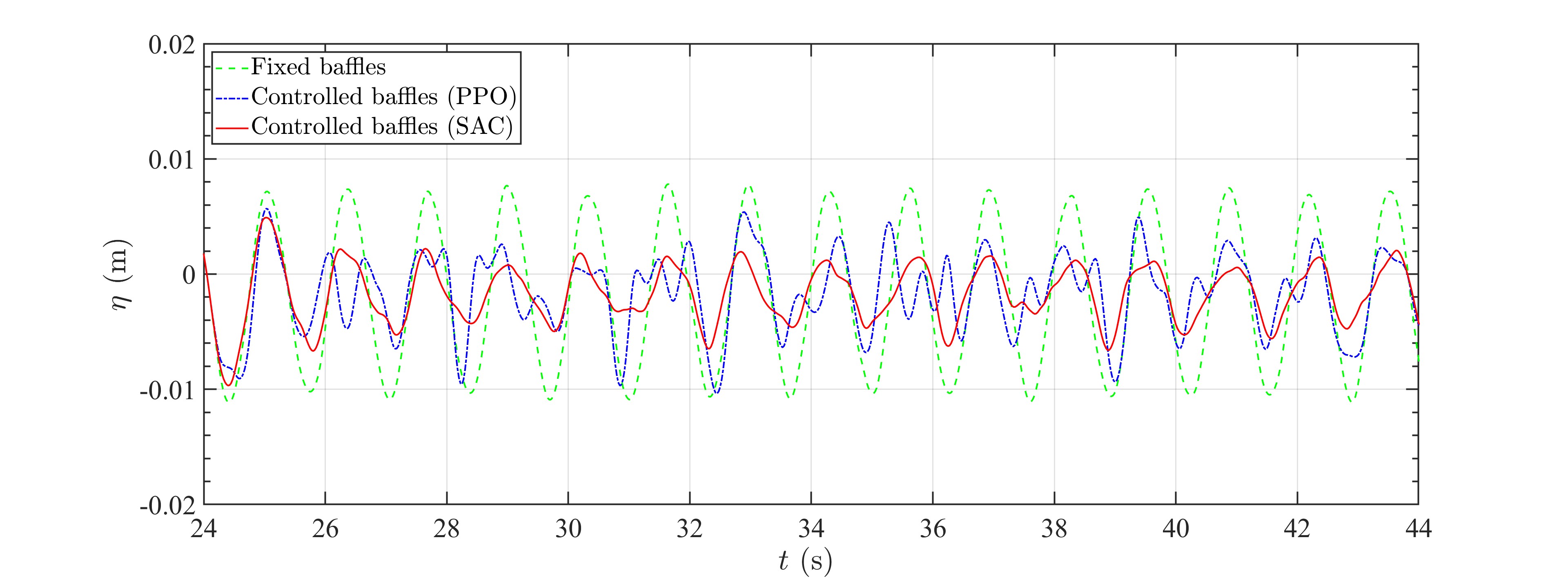}
\caption{The time evolution of the free surface height along the left wall under three conditions: fixed baffles and controlled baffles utilizing PPO and SAC.}\label{fig10}
\end{figure}

As illustrated in Figure \ref{fig10}, the active-controlled baffles significantly reduce the free surface height, effectively mitigating tank sloshing. Among the methods evaluated, the SAC algorithm exhibits superior performance, achieving a 68.81\% reduction in sloshing, a result comparable to that obtained by \cite{xie2021sloshing} using the TD3 algorithm, albeit slightly lower than the 81.48\% reduction achieved by TD3 Behavior Cloning (TD3BC). The primary reason for this discrepancy stems from differences in action definitions. In \cite{xie2021sloshing}, the baffle's velocity was directly used as the control action, resulting in sharp velocity fluctuations. Although this increases the y-direction displacement of the baffle, such rapid movements are impractical in real-world engineering applications and can introduce significant numerical errors during simulation.

\begin{figure}[htbp]
\centering
\includegraphics[width=\textwidth]{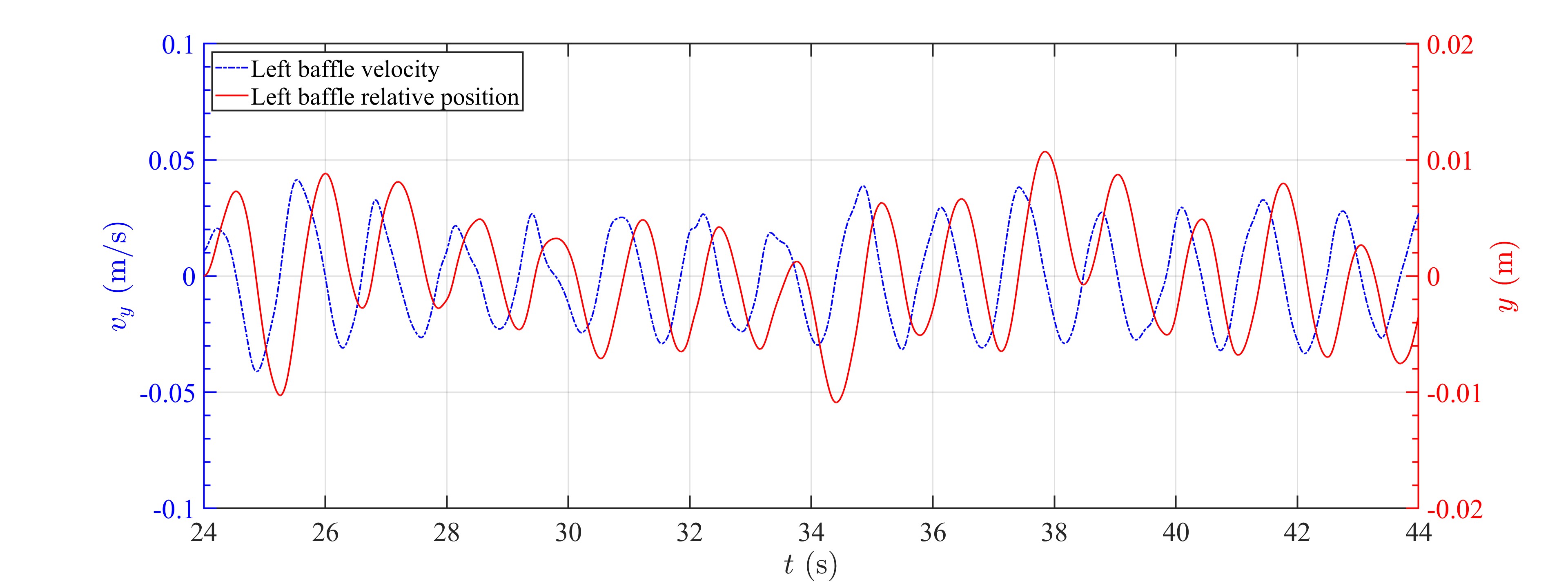}
\caption{The y-direction velocity and displacement relative to the initial position of the left baffle under the control of SAC.}\label{fig11}
\end{figure}

\begin{figure}[htbp]
\centering
\includegraphics[width=\textwidth]{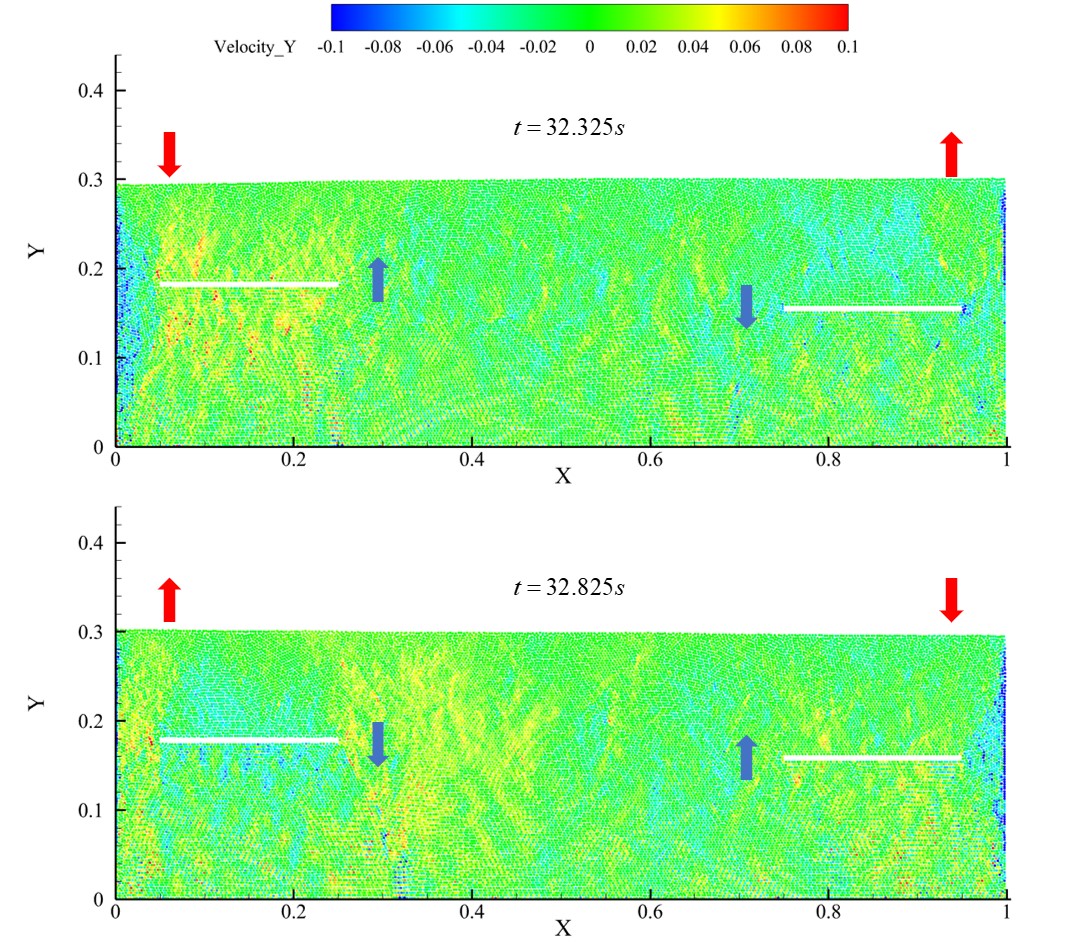}
\caption{The velocity contour plots in the y-direction at two representative moments. Due to the sloshing, the liquid surface above the left and right baffles exhibit opposing variations, resulting in opposite movement directions for the baffles. For instance, at $t$ = 32.325 s, the liquid surface on the left baffle drops from a peak to a trough, which should lead to negative velocity in the y-direction. However, the moving up baffle leads to significant changes in the y-direction velocity, generating positive velocity components.}\label{fig12}
\end{figure}

\begin{figure}[htbp]
\centering
\includegraphics[width=\textwidth]{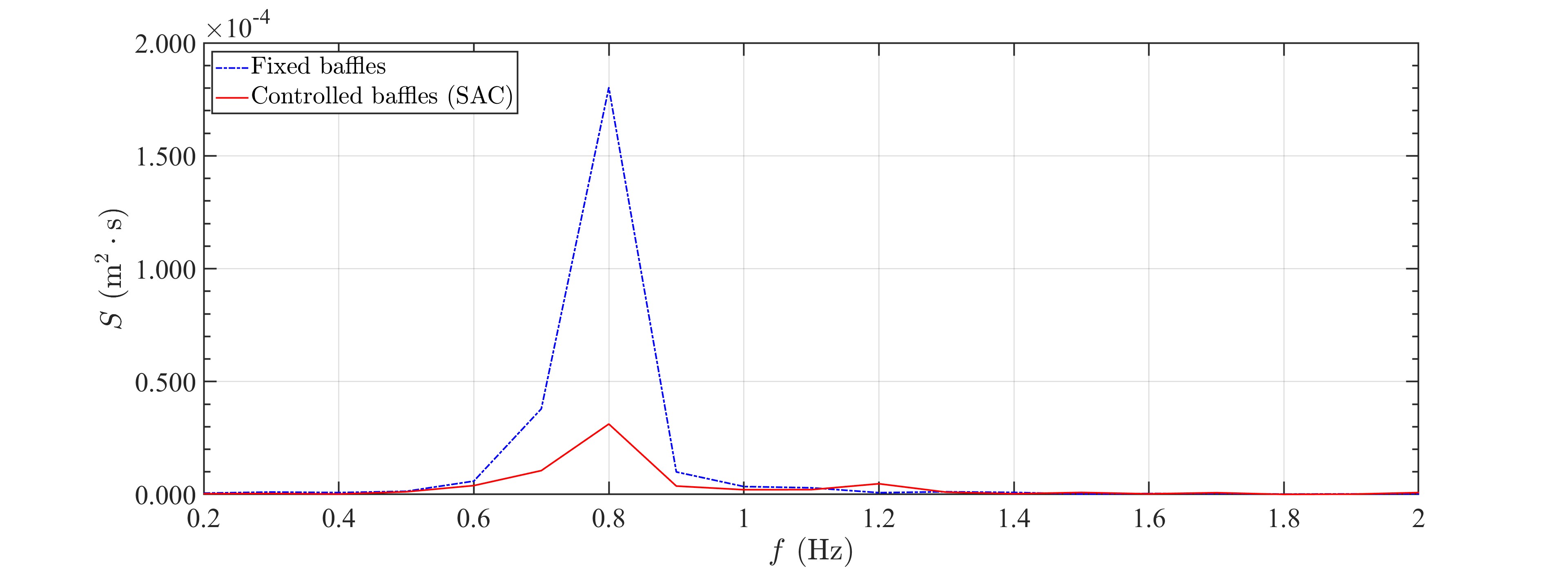}
\caption{The spectral analysis of the free surface elevation at the right wall of the tank under different control policies.}\label{fig13}
\end{figure}

In contrast, this study modifies the action output to represent changes in velocity, thereby addressing the aforementioned issues. As shown in Figure \ref{fig11}, the baffle's velocity and displacement curves exhibit apparent periodicity and correlate strongly with changes in the free surface height at the baffle location. Moreover, the contour plot in Figure \ref{fig12} demonstrates that the baffle's displacement consistently moves in the opposite direction of the free surface elevation, performing negative work on the liquid and thus reducing the kinetic energy, effectively dampening the sloshing. Additionally, the spectral analysis in Figure \ref{fig13} indicates that the y-direction movements of the baffles do not significantly change the tank's characteristic frequency, a finding that contrasts with the results reported by \cite{xie2021sloshing}.

\subsection{Case 2: Sloshing suppression with an elastic baffle}
\begin{figure}[htbp]
\centering
\includegraphics[width=0.6\textwidth]{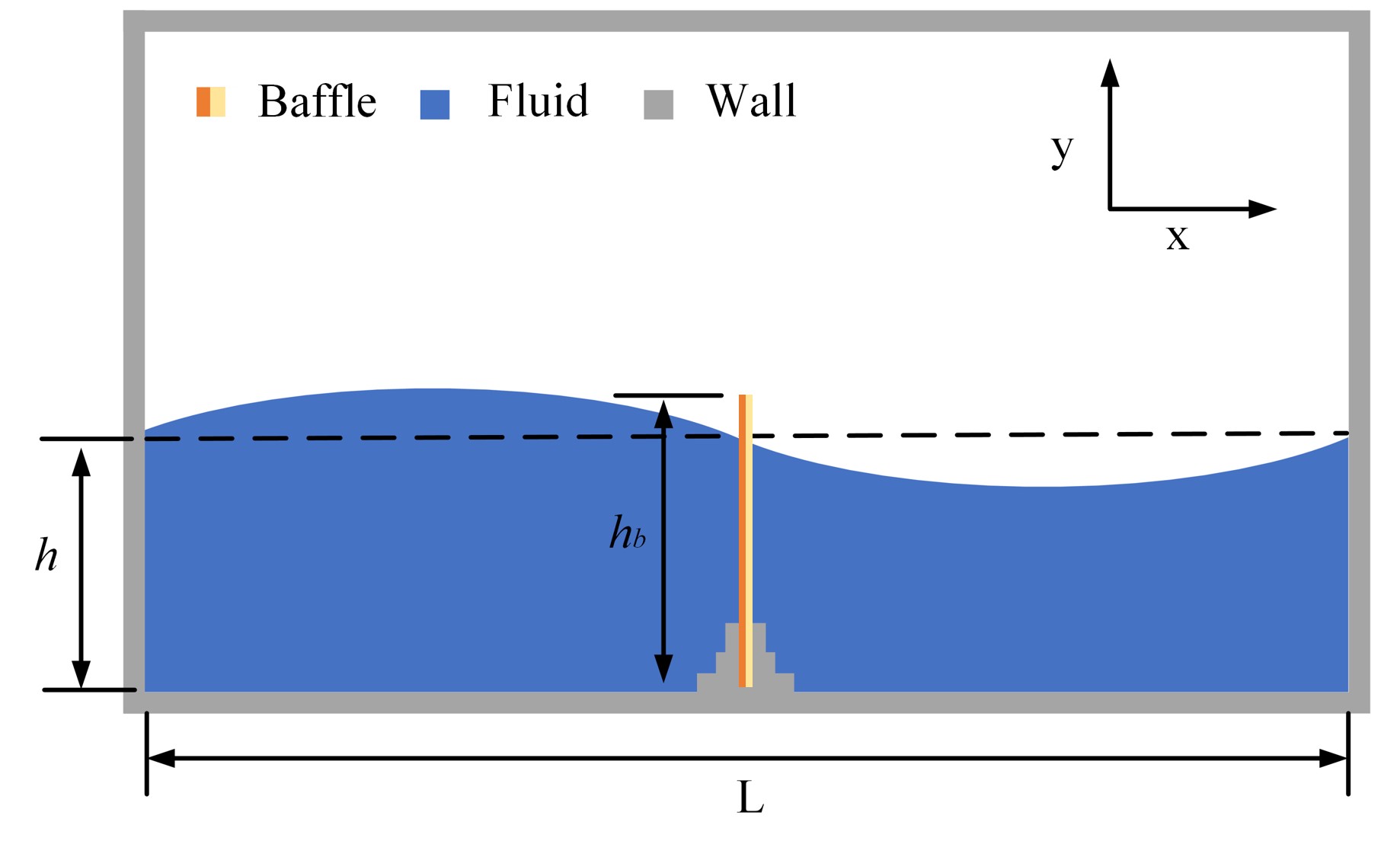}
\caption{The geometry of the tank includes a bottom-fixed elastic baffle, with the baffle positioned at the center of the tank.}\label{fig14}
\end{figure}

There is limited research on the problem of sloshing suppression using elastic baffles. \cite{ren2023experimental} conducted detailed experimental and 2D numerical studies with SPHinXsys \citep{ren2023numerical} on tank sloshing with elastic baffles, as shown in Figure \ref{fig14}. Their results demonstrated that elastic baffles can effectively reduce sloshing flow. They also investigate the effect of baffles with varying stiffness and different water depths $h$. However, the interaction between elastic plates and sloshing is mainly focused on passive deformation, with little research on the active control of elastic baffles, such as controlling the movement of the baffle or applying active strain to mitigate sloshing. Building on the work of \cite{ren2023numerical}, this section explores the effect of applying active strain, denoted as $\mathbb{E}_a$, to the elastic baffle to assess its effectiveness in mitigating sloshing. The specific form of the active strain in the x-direction, as outlined in \citet{curatolo2016modeling}, is given by
\begin{equation}
\label{eq:eq39}
\mathbb{E}_a = -\epsilon_0 \sin^2(\frac{\omega_b t + k_b Y + \psi}{2})h(Y)s(t),
\end{equation}
where $\epsilon_0 = 0.1$ represents the maximum shortening amplitude of the baffle, $\omega_b$ the angular frequency, which is equal to the excitation frequency $\omega_e$, wave number $k_b = (2\pi /\lambda)$ and $\lambda = 3 h_b$ is the wavelength. $\psi = \pi$ refers to the phase difference of the active strain on both sides of the baffle at the same height $Y$. The function $h(Y)$ describes the increasing shortening along the y-direction direction from the top to the bottom of the baffle, and $s(t)$ is introduced to ensure stability during the initial stage of the simulation
\begin{equation}
\label{eq:eq40}
\left\{
\begin{aligned}
h(Y) &= -\frac{Y^2 - h_b^2}{h_b^2} \\
s(t) & = 1 - \exp(-t/0.2).
\end{aligned}
\right.
\end{equation}

The tank employed in DRL training measures $L = 0.5$ m in length and has a water depth $h = 0.15$ m. The external excitation is applied with an amplitude $X = 0.01$ m and a frequency $\omega_e = 8.08$ rad/s. The baffle thickness is $l_b = 0.008$ m, with a height $h_b = 0.2$ m, and is constrained by a simplified bottom slot with 0.026 m. The material properties of the baffle include Young's modulus $E_b = 30$ MPa, density $\rho_b = 1250\, \text{kg/m}^3$, and Poisson's ratio $\nu_b = 0.47$.

The state $s_n$, similar to \emph{Case 1}, monitors the free surface height and velocity. Given that the baffle undergoes deformation, four additional monitoring points are added on each side of the baffle to observe the active strain and position. The action $a_n$ primarily controls $\Delta \epsilon_0$, with the constraint $|\Delta \epsilon_0| \leq 0.025$ in one action time step 0.1 s. For the reward $r_n$, the parameter $H$ is set to 0.05 m, and only the penalty term \( p_0 \) is retained in Eq.~(\ref{eq:eq36}), as shown below
\begin{equation}
p_0 = 
\begin{cases} 
-1, & \text{if } \epsilon_0 < 0 \text{ or } \epsilon_0 > 0.2 , \\
0, & \text{otherwise}.
\end{cases}
\label{eq:eq41}
\end{equation}

The SAC algorithm is used for training, keeping the hyperparameters consistent with those in \emph{Case 1}. Each training episode begins at \(0.5 \ \text{s}\), executing 200 actions over 20 seconds.

\subsubsection{Numerical model validation}
\begin{figure}[htbp]
\centering
\includegraphics[width=\textwidth]{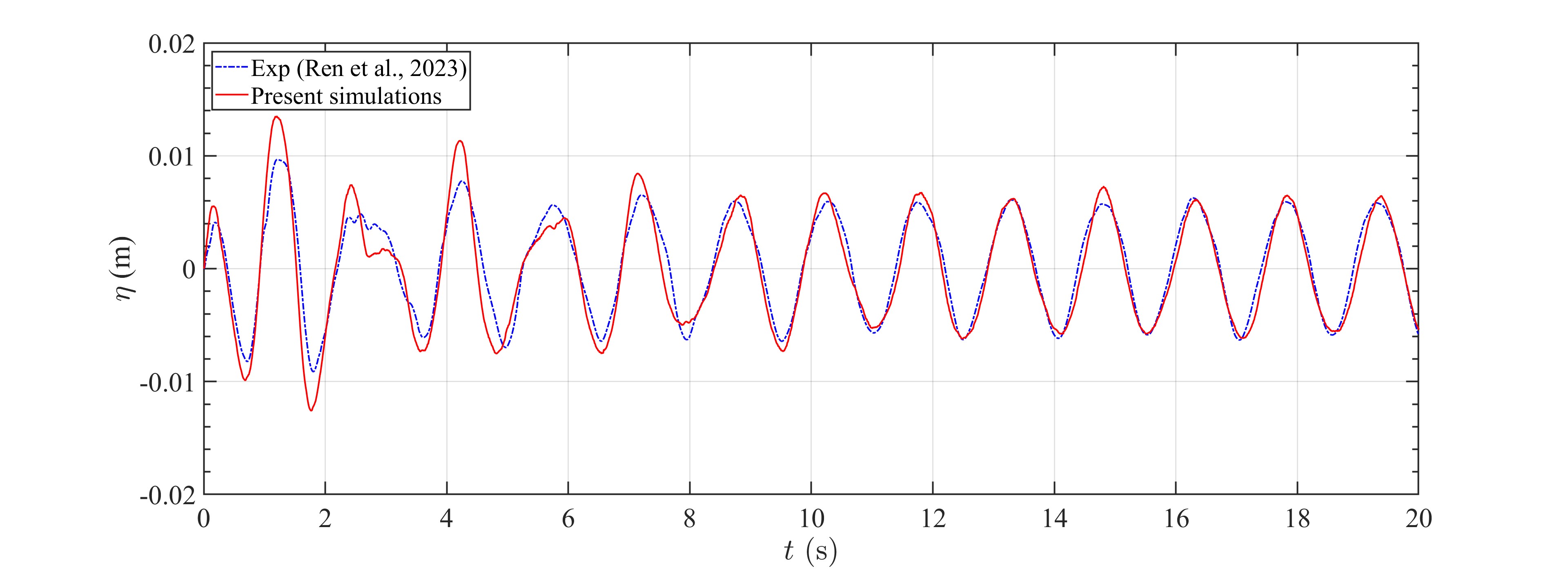}
\caption{The comparison of free surface height between the numerical simulation and the experiment at the left wall.}\label{fig15}
\end{figure}

In the experiment, the tank length is $L$ = 1.0 m, and the baffle height is $h_b$ = 0.2 m. The system is subjected to an external excitation with a frequency of 4.14 rad/s and an amplitude of 0.01 m. From Figure \ref{fig15}, we can see that our numerical model can capture the free surface height very well with a particle resolution of $dp = 0.002 \ \text{m}$.

\subsubsection{Results}
\begin{figure}[htbp]
\centering
\includegraphics[width=\textwidth]{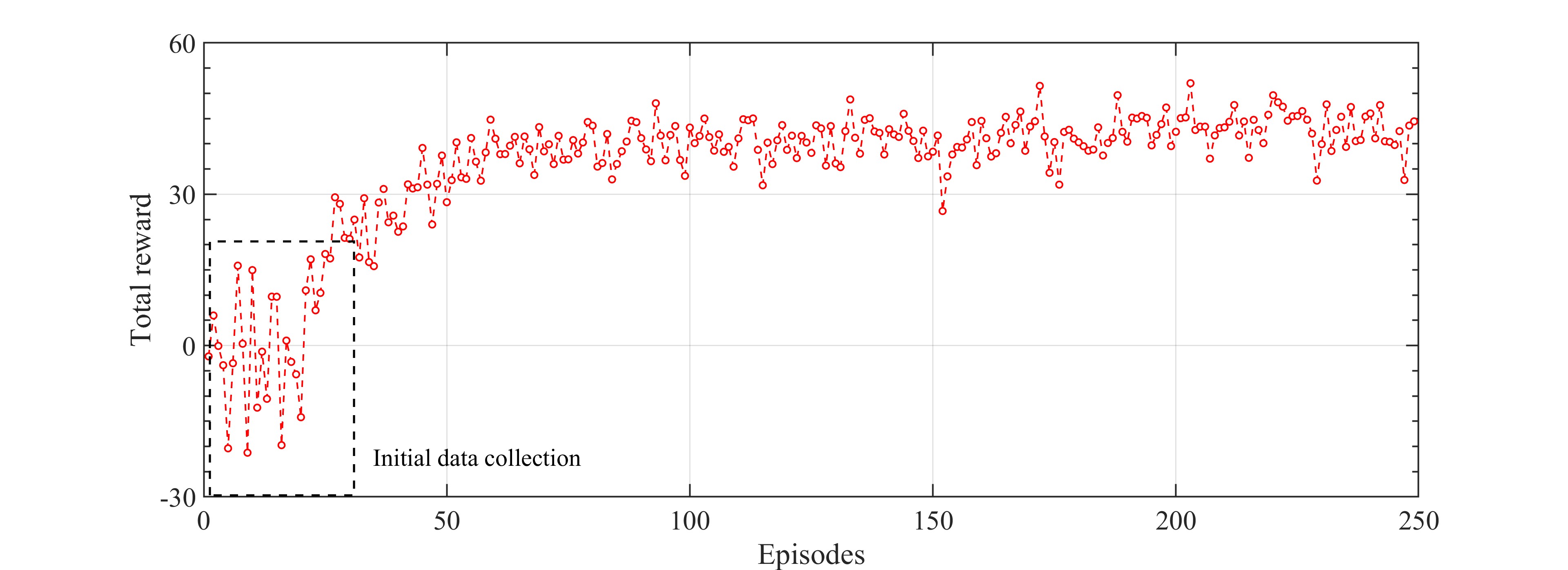}
\caption{The training curves obtained from training the elastic baffle using the SAC algorithm.}\label{fig16}
\end{figure}
\begin{figure}[htbp]
    \centering
    \begin{subfigure}{\textwidth}
        \centering
        \includegraphics[width=\textwidth]{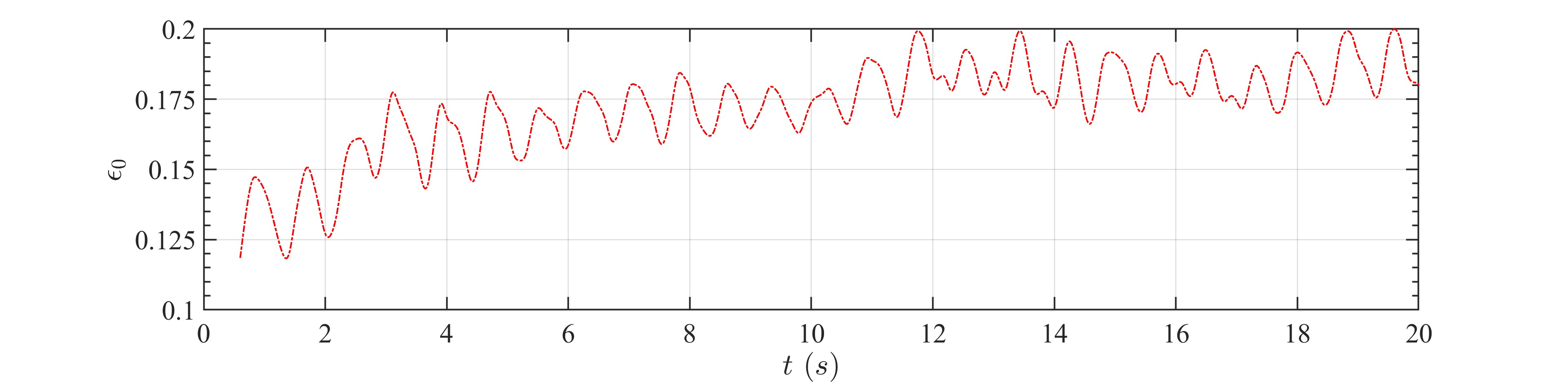}
        \caption{Amplitude of the active strain.}
    \end{subfigure}
    
    \begin{subfigure}{\textwidth}
        \centering
        \includegraphics[width=\textwidth]{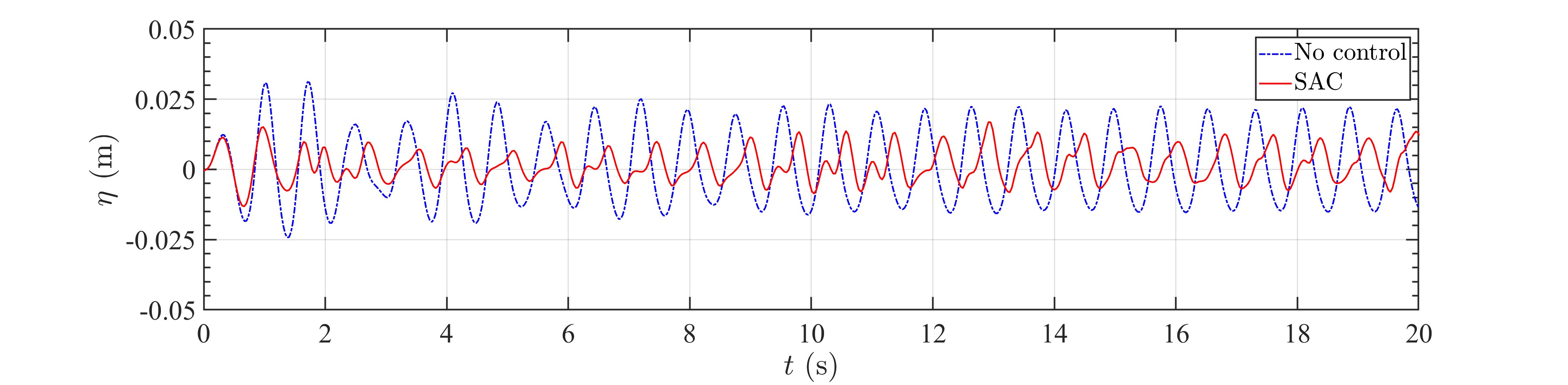}
        \caption{Free surface height at the left wall.}
    \end{subfigure}
    
    \begin{subfigure}{\textwidth}
        \centering
        \includegraphics[width=\textwidth]{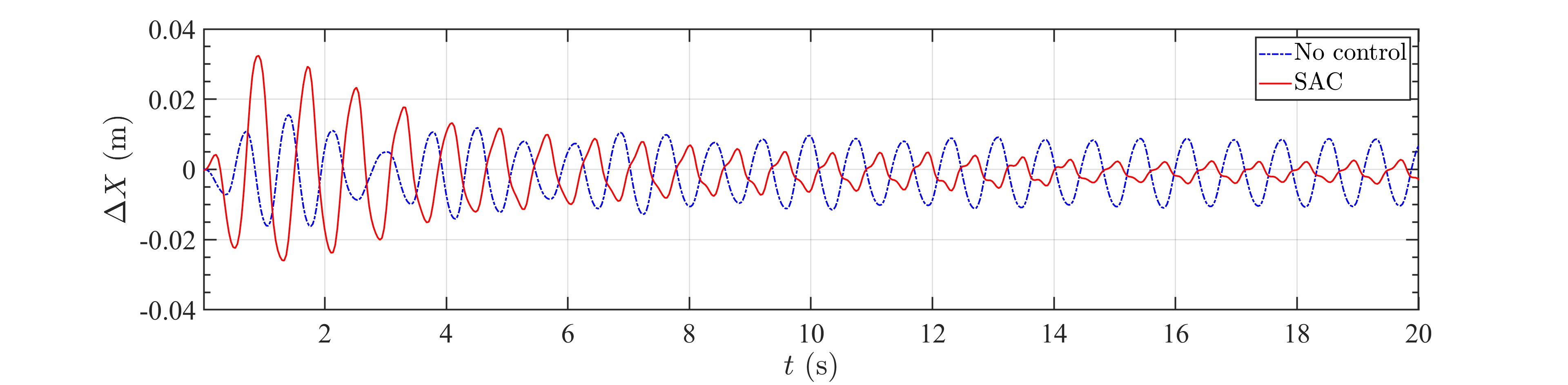}
        \caption{The deformation of the top of the elastic plate in the x-direction compared to its initial position.}
    \end{subfigure}

    \begin{subfigure}{\textwidth}
        \centering
        \includegraphics[width=\textwidth]{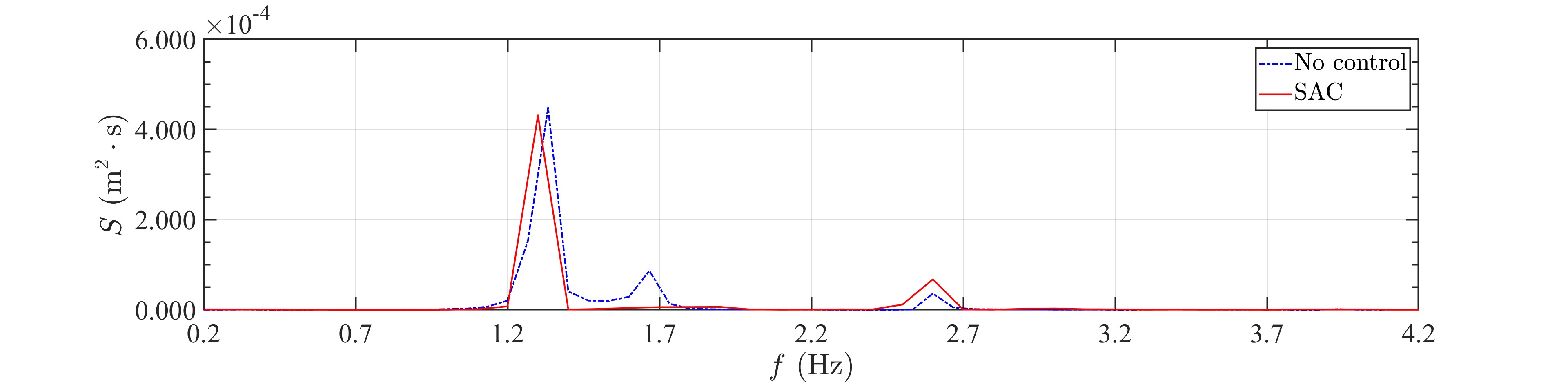}
        \caption{The spectral analysis of the free surface elevation at the left wall.}
    \end{subfigure}
    
    \caption{The amplitude of the active strain controlled with SAC (a), and its effects on the free surface height (b), the deformation of the baffle (c), and the spectral analysis (d).}
    \label{fig17}
\end{figure}

Figure \ref{fig16} shows that after 60 episodes, the SAC algorithm discovers an improved strategy. Figure \ref{fig17} (a) depicts the active strain's amplitude variation. It can be seen that during the first 10 seconds, the amplitude exhibits an overall increasing trend, and after 10 seconds, it stabilizes, fluctuating around 0.18. Notably, applying active strain significantly affects the overall free surface height within the tank. As shown in Figure \ref{fig17} (b), the sloshing at the left wall is substantially suppressed, with a reduction of approximately 38.7\% after stabilization. The contour plots clearly demonstrate that the active strain causes the bending direction of the elastic baffle to be exactly opposite to the sloshing direction. This results in the elastic plate performing negative work on the fluid, thereby suppressing the sloshing. Furthermore, Figure \ref{fig17} (c) shows that the active strain changes the deformation period of the baffle, with the phase difference being about half a period after stabilization. In addition, the nonlinear changes of the elastic plate also alter the sloshing frequency of the liquid in the tank, as shown in Figure \ref{fig17} (d). A frequency of approximately 1.7 Hz did not appear in the calculations controlled by the SAC algorithm.
\begin{figure}[htbp]
\centering
\includegraphics[width=0.6\textwidth]{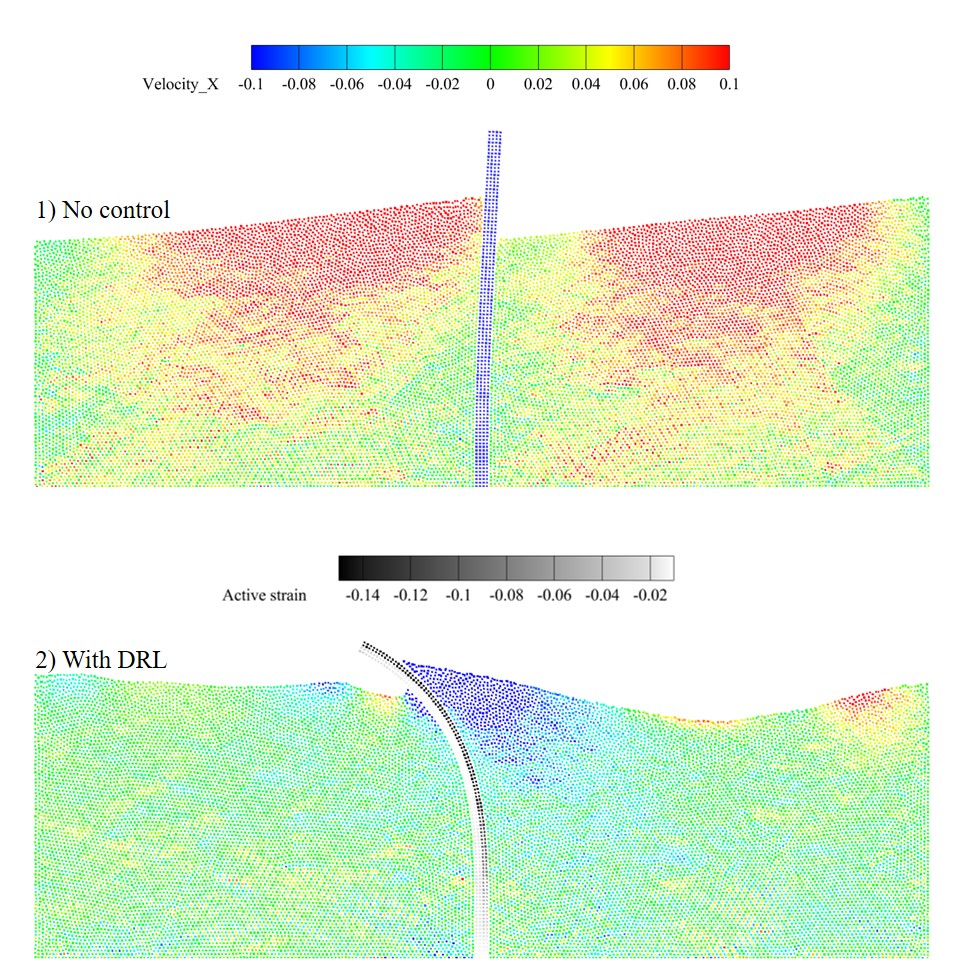}
\caption{The x-direction velocity field of the tank, no control and with SAC \( (t = 7.45 \ \text{s} )\).}\label{fig18}
\end{figure}

\subsection{Case 3: Wave energy capture optimization of an OWSC}
\begin{figure}[htbp]
\centering
\includegraphics[width=0.6\textwidth]{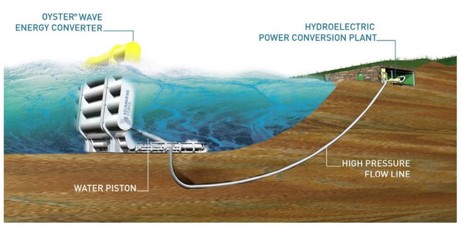}
\caption{The Oyster\textsuperscript{\textregistered} developed by Aquamarine Power Ltd.}\label{fig19}
\end{figure}

Since the beginning of the 21st century, the potential for harnessing wave energy has become increasingly viable \citep{folley2009analysis}. Among the most common wave energy converters (WECs) utilized in nearshore waters are OWSCs, which typically feature bottom-hinged flap mechanisms. A notable commercial example of this technology is the Oyster \citep{folley2007design}. The structure of the Oyster, depicted in Figure \ref{fig19}, includes a flap whose upper edge extends above the water surface \citep{cheng2019fully}. The flap is also attached to the base by a hinge and oscillates in response to incident waves. The resulting oscillatory motion drives a hydraulic pump to pressurize water and transfer it through a pipeline to a hydroelectric turbine, generating electricity \citep{renzi2014does}.

\begin{figure}[htbp]
\centering
\includegraphics[width=\textwidth]{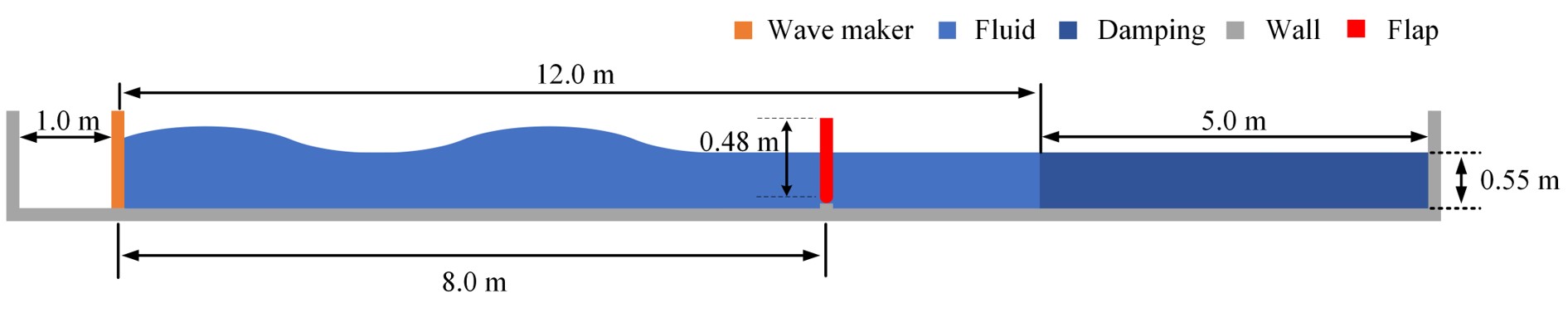}
\caption{The structure of the wave tank and the OWSC in 2D simulations.}\label{fig20}
\end{figure}

Numerical studies on OWSCs have been extensively conducted \citep{wei2015wave, schmitt2016optimising}, and 2D simplified modeling has been demonstrated as a practical approach for accelerating shape optimization while maintaining reasonable accuracy \citep{zhang2021efficient}. The 2D structure used in this section is illustrated in Figure \ref{fig20}. The simplified representation of the power take-off (PTO) system of the OWSC contains the base, flap, and hinge. The base o has a height of 0.1 m while the flap has a height of 0.48 m and a width of 0.12 m. It is positioned 8.0 m away from the wave maker and is connected to the base with a damped hinge. The center of the hinge is located 0.06 m above the base, and the hinge can be directly controlled with the damping coefficient $k_d$.

A piston-type wave maker consisting of a group of dummy particles generates the second-order Stokes wave \citep{zhang2022smoothed}. The linear wave make theory by \cite{madsen1971generation} is adopted, where the wave maker motion $r_{m}$ is 
\begin{equation}
\label{eq:eq42}
r_{m} = -S_{0}\cos(2\pi f t) - S_{0}(\frac{3H\sin(4\pi f t)}{4n_{0}h(4\sinh^{2}(kh) - n_{0} / 2)}).
\end{equation}
Here $S_{0} = H n_{0} / (2 \tanh(kh)$), $n_{0} = (\sinh(2kh) + 2kh) / (2\sinh(2kh))$, $H$ the wave height, $k$ is the wave number followed by the dispersion relation \citep{madsen1971generation}
\begin{equation}
\label{eq:eq43}
\omega_c^2 = g k\tanh(kh),
\end{equation}
where $\omega_c = 2\pi f$ is the wave angular frequency.

Besides, the damping zone is set to mitigate the impact of wave reflection off the wall on the motion of the OWSC. The velocity of the particle $\mathbf{v}$ is then given by
\begin{equation}
\label{eq:eq44}
\mathbf{v} = \mathbf{v}_{0}(1.0 - \alpha \Delta t_{ac} (\frac{\mathbf{r} - \mathbf{r}_{0}}{\mathbf{r}_{1} - \mathbf{r}_{0}})).
\end{equation}
$\mathbf{v}_{0}$ the fluid particle velocity at the entrance of the damping zone, the reduction coefficient $\alpha$ is set as 5.0. $\mathbf{r}_{0}$ and $\mathbf{r}_{1}$ are the initial and final position vectors of the damping zone.

At the end of each time step $\Delta t_{ac}$, the total force $\mathbf{F}$ and total torque $\tau$ calculated from SPHinXsys are transmitted to Simbody for solving the Newton-Euler equations
\begin{equation}
\left\{
\begin{aligned}
\label{eq:eq45}
\mathbf{F} &= \sum_{a\in N}(\mathbf{f}_{ap} + \mathbf{f}_{a\nu}) = m\mathbf{I}_{0}\frac{d\mathbf{v}}{dt} \\
\tau &= \sum_{a\in N}(\mathbf{r}_{a} - \mathbf{r}_{g}) \times (\mathbf{f}_{ap} + \mathbf{f}_{a\nu}) = \mathbf{J}_{0}\frac{d\Omega}{dt} - k_{d}\Omega.
\end{aligned}
\right.
\end{equation}
Here, $N$ denotes the total number of structure particles. $\mathbf{f}_{ap}$ and $\mathbf{f}_{a\nu}$ are the pressure and viscous force on the structure, which can be directly obtained from Equation \ref{eq:eq20}. $m$ is the mass of the flap, $\mathbf{I}_{0}$ the identity matrix. $\mathbf{r}_{g}$ is the position vector of the flap mass center, $\mathbf{J}_{0}$ the moment of the inertia about the center of mass, $\Omega$ the angular velocity of the flap.

\begin{figure}[htbp]
\centering
\includegraphics[width=\textwidth]{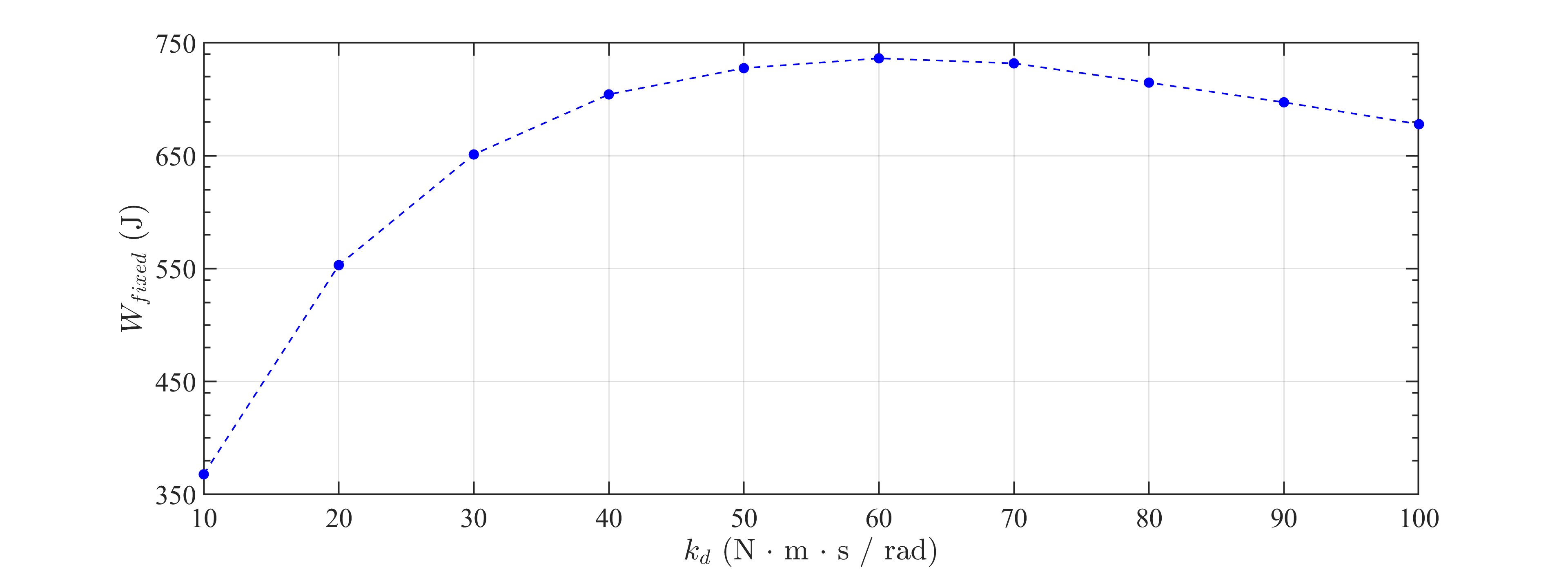}
\caption{The variations of the total wave energy conversion $W_{fixed}$ in terms of damping coefficients.}\label{fig21}
\end{figure}

The state $s_n$, similar to \emph{Case 2}, monitors the free surface height and velocity. The position and the angular velocity of the flap are also considered. The action $a_n$ primarily controls $\Delta k_d$, with the constraint $|\Delta k_d| \leq 25 \, \text{N} \cdot \text{m} \cdot \text{s} / \text{rad}$, which can be rapidly implemented through the damping update definition in Simbody. The instance energy capture in one action time step 0.1 s is defined with \citep{senol2019enhancing}
\begin{equation}
\label{eq:eq46}
P_{out} = \sum_{n=0}^{M - 1}k_{d}^{n}(\frac{\Omega_{n+1} + \Omega_{n}}{2})^{2}, 
\end{equation}
where $M = 10$. The reward can be calculated with
\begin{equation}
\label{eq:eq47}
r_{n} = p_{e} - p_{b} + p_0.
\end{equation}
Here, $P_{out}$ represents the instantaneous energy capture with $k_d = 60 \text{N} \cdot \text{m} \cdot \text{s} / \text{rad}$, which corresponds to the optimal solution under the fixed damping condition, as shown in Figure \ref{fig21}. The value of $p_0$ will be set to $-1$ if $k_d < 0$ or $k_d > 100\ \text{N} \cdot \text{m} \cdot \text{s} / \text{rad}$. 

The SAC algorithm is used for training, with the hyperparameters kept consistent with those in \emph{Case 1}. The training collector starts at 4.0 s in each episode and runs for 200 actions.

\subsubsection{Numerical model validation}
\begin{figure}[htbp]
\centering
\includegraphics[width=\textwidth]{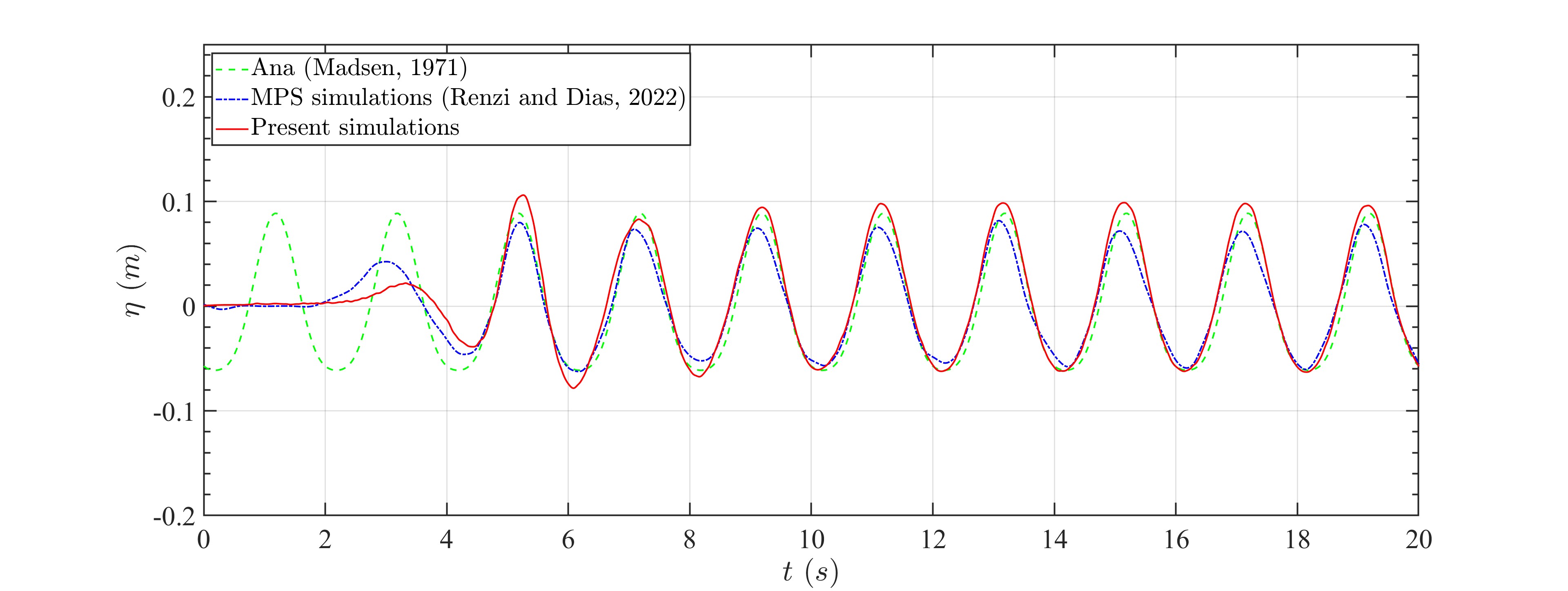}
\caption{Comparisons of free surface height at $x$ = 6.0 m with the analytical and simulation results.}\label{fig22}
\end{figure}
We validated the wave generation in the absence of the OWSC, as shown in Figure \ref{fig22}. The flume dimensions are \(L = 15\, \text{m}\) and \(h = 0.64\, \text{m}\), and the wave parameters are \(h = 0.15\, \text{m}\) and \(f = 0.5\, \text{Hz}\) \citep{madsen1971generation, renzi2022application}. We can see that our numerical result is in excellent agreement with the theoretical solution. The validation of wave interaction with the OWSC can be obtained from our previous work \citep{zhang2021efficient}.

\subsubsection{Results}
\begin{figure}[htbp]
\centering
\includegraphics[width=\textwidth]{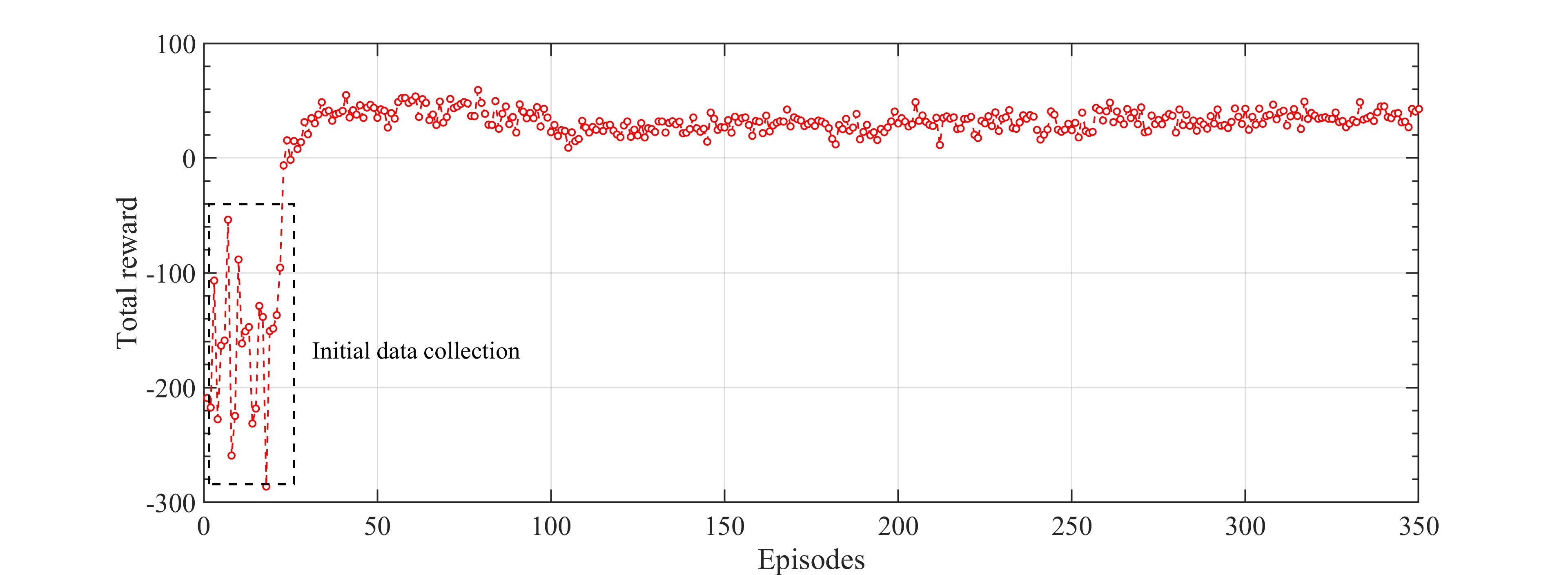}
\caption{The training curves obtained from training the OWSC using the SAC algorithm.}\label{fig23}
\end{figure}
\begin{figure}[htbp]
    \centering
    \begin{subfigure}{\textwidth}
        \centering
        \includegraphics[width=\textwidth]{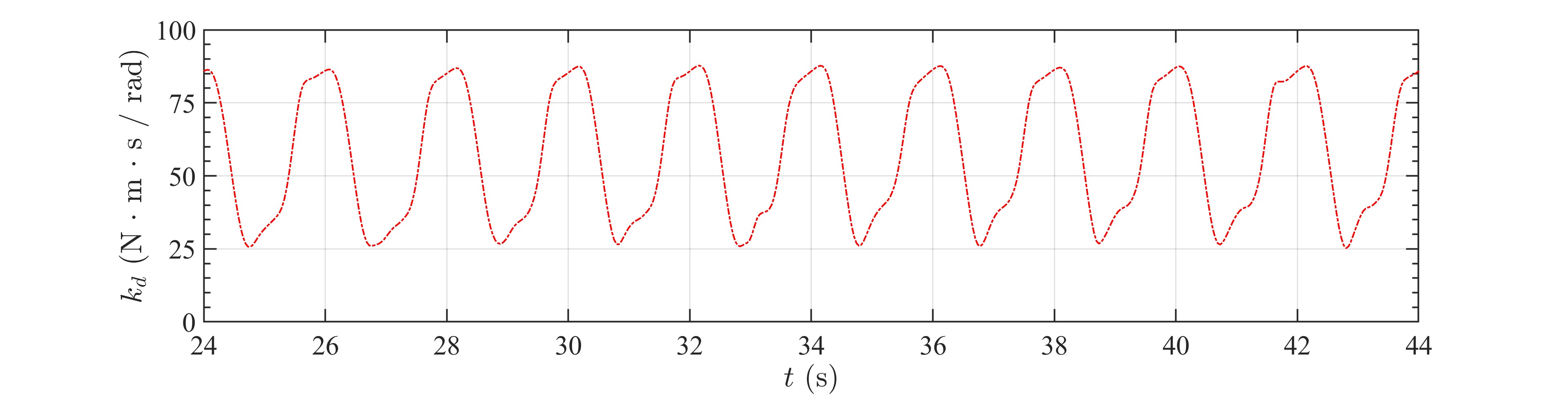}
        \caption{Damping coefficient of the PTO system.}
    \end{subfigure}
    
    \begin{subfigure}{\textwidth}
        \centering
        \includegraphics[width=\textwidth]{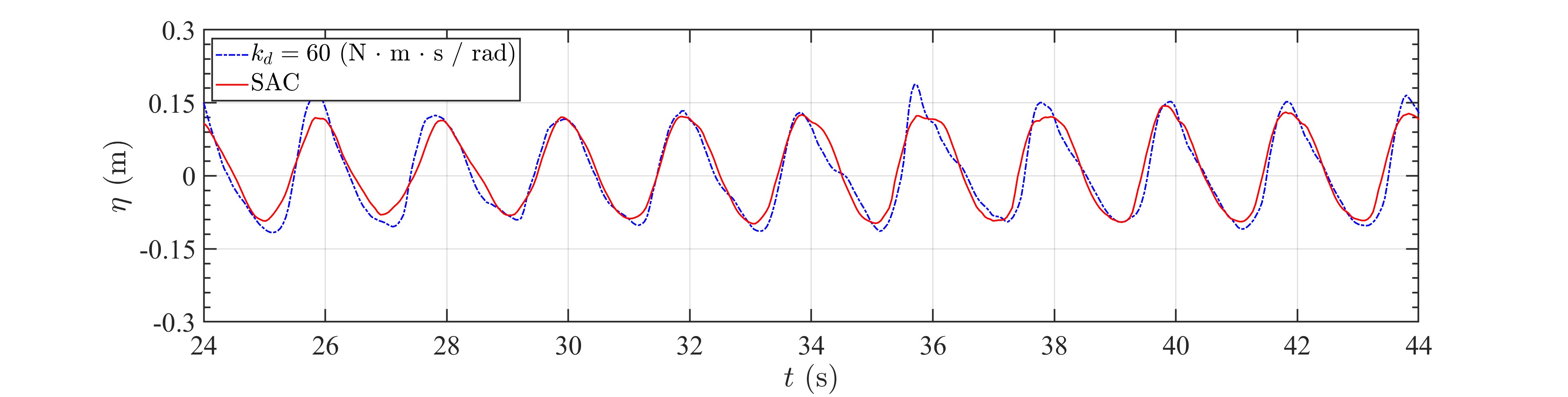}
        \caption{Free surface height at $x$ = 7.5 m in front of the flap.}
    \end{subfigure}
    
    \begin{subfigure}{\textwidth}
        \centering
        \includegraphics[width=\textwidth]{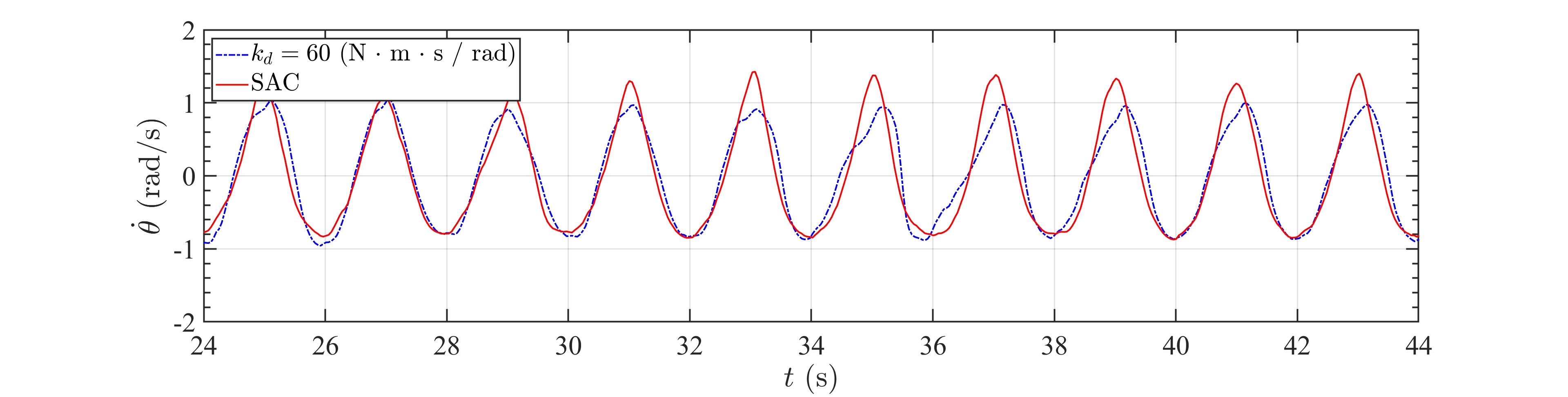}
        \caption{Angular velocity of the flap.}
    \end{subfigure}

    \begin{subfigure}{\textwidth}
        \centering
        \includegraphics[width=\textwidth]{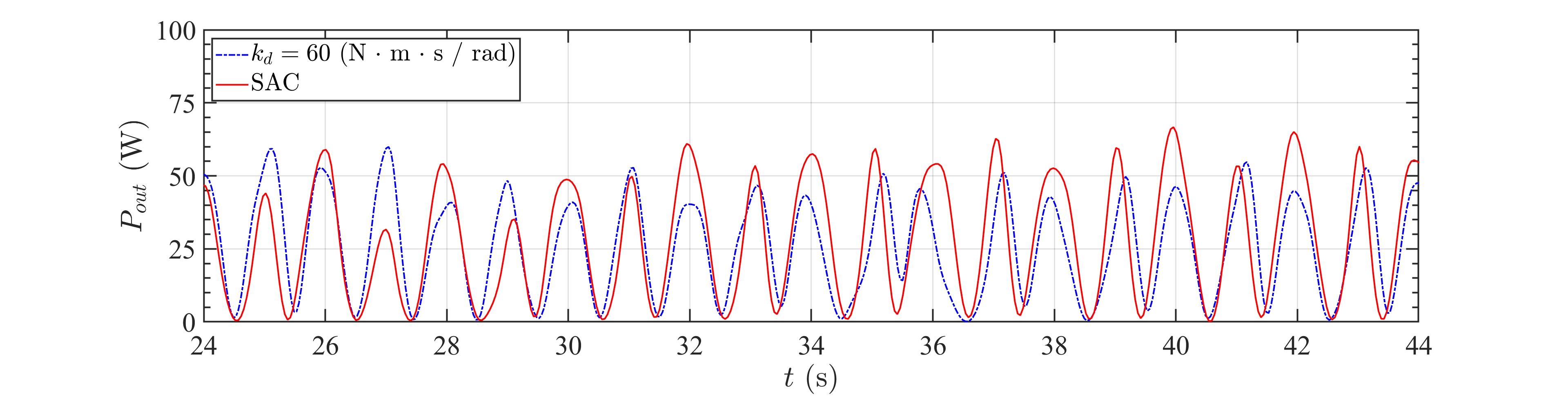}
        \caption{Instantaneous power capture.}
    \end{subfigure}
    
    \caption{The damping coefficient of the PTO system controlled with SAC (a), and its effects on the free surface height (b), the angular velocity of the flap (c), and instantaneous power capture (d).}
    \label{fig24}
\end{figure}
\begin{figure}[htbp]
\centering
\includegraphics[width=\textwidth]{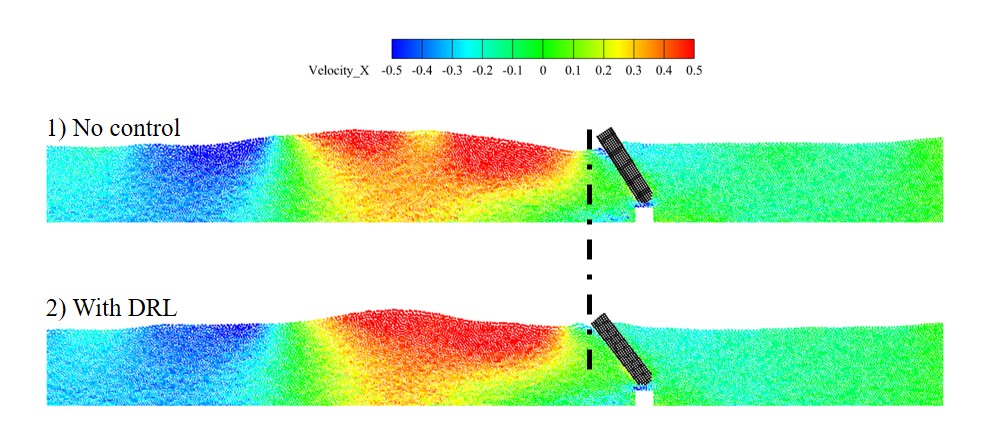}
\caption{The x-direction velocity field before and after the OWSC, no control and with SAC  \( (t = 43.4 \ \text{s} )\).}\label{fig25}
\end{figure}

The training curve in Figure \ref{fig23} shows that the SAC algorithm can find an optimal strategy after approximately 50 episodes. Analyzing the period from 24 s to 44 s, Figure \ref{fig25} shows that the periodic characteristics of the free surface wave height at the OWSC’s flap ($x = 7.5$ m) align closely with the damping coefficient of the PTO system controlled by the SAC algorithm. Taking a wave period between 37.4 s and 39.4 s as an example, when the wave crest passes through the OWSC, the damping coefficient increases, reaching a maximum of 87 $\text{N}\cdot\text{m}\cdot\text{s}/\text{rad}$. Due to the high energy density in the crest segment, the angular velocity of the flap remains almost unchanged despite the increased damping, enhancing energy capture. The wave energy density is lower in the trough segment, but the damping coefficient significantly decreases to around 25 $\text{N} \cdot \text{m} \cdot \text{s} / \text{rad}$. As the damping is reduced, the angular velocity of the flap increases noticeably, slightly boosting energy capture, as confirmed by the instantaneous power of the PTO system shown in Figure \ref{fig24} (d). Furthermore, the contour plot in Figure \ref{fig25} indicates that the increased angular velocity of the flap during the trough segment causes a shift in the maximum deflection angle of the flap, which moves approximately 0.18° to the left. 

Overall, from 24 to 44 seconds, the total energy captured with a fixed damping coefficient is about 514.38 J, while the energy captured using the SAC-controlled system is 556.82 J, representing an overall efficiency improvement of 8.25\%. However, it is important to note that this study is based on 2D numerical simulations and does not consider the impact of waves on the OWSC in a 3D environment, which requires further exploration in future studies.

\subsection{Case 4: Training of muscle-driven fish swimming in the vortexs}
\begin{figure}[htbp]
\centering
\includegraphics[width=\textwidth]{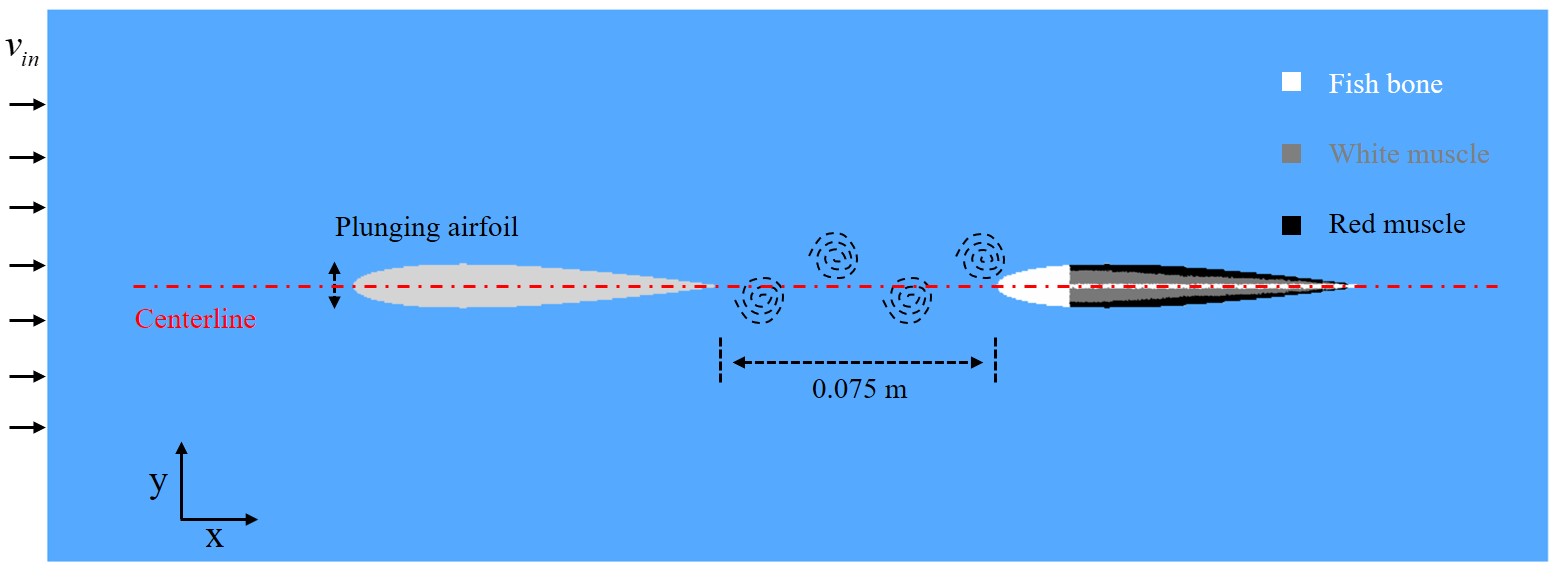}
\caption{The muscle-driven fish swimming behind a plunging airfoil. Both the airfoil length $L_1$ and the fish length $L_f$ use the NACA0012 shape, with a chord length of 0.1 m. The initial distance between the airfoil and the fish is 0.075 m, and the up-and-down boundaries are free-stream boundaries. The fluid domain has dimensions of \( 0.5\,\text{m} \times 0.15\,\text{m} \), and the inlet velocity $v_{in}$ is set to 0.2\,\text{m/s}.}\label{fig26}
\end{figure}

\cite{thandiackal2023line} conducted an experimental study on fish in-line swimming dynamics. They used rainbow trout as the model organism and placed a NACA0012 airfoil in front of the fish. They simulated the vortex street generated by the fish's swimming by applying plunging and pitching motions to the airfoil. The results indicated that the pressure on its head decreases when the fish swims behind the airfoil, improving swimming efficiency. Currently, there are few related numerical studies about muscle-driven fish in-line swimming. \cite{zhao2023hydrodynamic} conducted a numerical study on the collective swimming behavior of two self-propelled biomimetic fish using PID control. In this section, we preliminarily explore the possibility of using DRL to control a muscle-driven fish to swim steadily along the centerline behind the airfoil in a 2D numerical environment, as shown in Figure \ref{fig26}.

Considering the high Reynolds number ($Re \approx 10000$) of fish swimming, the transport-velocity formulation is adopted to mitigate issues such as particle clumping or void regions in numerical simulations with the WCSPH method \citep{zhang2020dual} in the form of
\begin{equation}
\label{eq:eq48}
\frac{d\Bar{\mathbf{v}_i}}{dt} = \frac{d\mathbf{v}_i}{dt} - \frac{2}{\rho_i}\sum_{j} p^0 \nabla W_{ij}V_{j},
\end{equation}
where $p^0$ is the background pressure and $\Bar{\mathbf{v}_i}$ represents the particle transport velocity.

Free-stream boundary conditions \citep{zhang2023lagrangian} were also applied to minimize computational cost and reduce the impact of walls on the fish swimming within the region. The discrete of position $\mathbf{r}_i$ can be expressed as
\begin{equation}
\label{eq:eq49}
\nabla \cdot \mathbf{r}_{i} = \sum_{j} \mathbf{r}_{ij} \cdot \nabla W_{ij}V_{j}.
\end{equation}
This equation determines the appropriate threshold $\gamma$, where $\nabla \cdot \mathbf{r}_{i} < \gamma = 0.75d$ indicates that the particles are surface particles. Here, $d$ is the dimension. Additionally, the spatio-temporal identification method assists in accurately identifying the properties of inner particles in low-pressure regions \citep{zhang2023efficient}. The density and velocity of all surface particles are corrected with
\begin{equation}
\left\{
\begin{aligned}
\label{eq:eq50}
\rho_i^{f} &= \rho_i^{n} + \max(0, \rho_i^{f} - \rho_i^{n})\frac{\rho_0}{\rho_i^{f}} \\
v_{x} &= v_{x} + (v_{in} - v_{x})\frac{\min(\rho_i^{f}, \rho^0)}{\rho^0}.
\end{aligned}
\right.
\end{equation}

The airfoil is treated as a rigid body, and its motion is simplified to pure plunging. The equation of motion in y-direction is given as
\begin{equation}
\label{eq:eq51}
r_{y} = S_1\sin(2\pi f_1 t),
\end{equation}
where $S_1 = 0.003$ m is the plunging amplitude, $f_1 = 4.0$ Hz the plunging frequency. Under this configuration, the vortices generated by the airfoil are very similar in structure and size to those produced by the swimming fish, making it suitable for describing in-line swimming in fish schools.
\begin{table}[htbp]
    \centering
    \caption{Basic material properties of the fish body.}
    \label{table2}
    \begin{tabular}{@{}lccc@{}}
        \toprule
        Properties & Young's modulus (MPa) & Density ($\text{kg/m}^3$) & Poisson's ratio \\ \midrule
        Bones & \(1.1 \times 10^6\) & 1050 & 0.49 \\
        White muscles & \(0.5 \times 10^6\) & 1050 & 0.49 \\ 
        Red muscles & \(0.8 \times 10^6\) & 1050 & 0.49 \\ \bottomrule
    \end{tabular}
\end{table}

The fish body consists of three main parts: the bones, white muscles, and red muscles \citep{curatolo2016modeling}. The basic material properties are shown in Table \ref{table2}. During cruising, the fish relies on red muscle for propulsion \citep{jayne1993red}, and the active strain is mainly applied to the red muscle, following the form of Equation (\ref{eq:eq39})
\begin{equation}
\label{eq:eq52}
\left\{
\begin{aligned}
\mathbb{E}_f &= -\epsilon_0 \sin^2(\frac{\omega_f t + k_f (L_f - X) + \psi}{2})h(X)s(t) \\
h(X) &= \frac{X^2}{L_f^2},
\end{aligned}
\right.
\end{equation}
where $\epsilon_0 = 0.12$, $\omega_f = 2 \pi f_1$, $k_f = 2 \pi$, and $\psi = \pi$.

The work of \cite{gunnarson2021learning} indicates that if the state \(s_n\) includes velocity information from the vortices, the agent's learning efficiency can be significantly improved. Since fish use lateral lines to sense changes in water flow, we distributed five measurement probes on each side of the fish's muscles to measure the water velocity and pressure.

The action \( a_n \) controls the variation of \(\epsilon_0\). When the amplitude of the red muscle on the left side of the fish increases, such that \(\epsilon_l = \epsilon_l + \Delta \epsilon_0\), the amplitude of the right-side muscle correspondingly decreases, \(\epsilon_r = \epsilon_r - \Delta \epsilon_0\), in order to control the fish’s swimming direction \citep{zhao2023hydrodynamic}. In one action time step 0.025 s, the constraint is \( |\Delta \epsilon_0| \leq 0.005 \). SAC is chosen to train 400 actions in one episode.

The definition of the reward \( r_n = 1.0 - |\Delta Y| \) is straightforward, which is related to the distance between the average position of the fish and the centerline in the y-direction.

\subsubsection{Numerical model validation}
\begin{figure}[htbp]
\centering
\includegraphics[width=\textwidth]{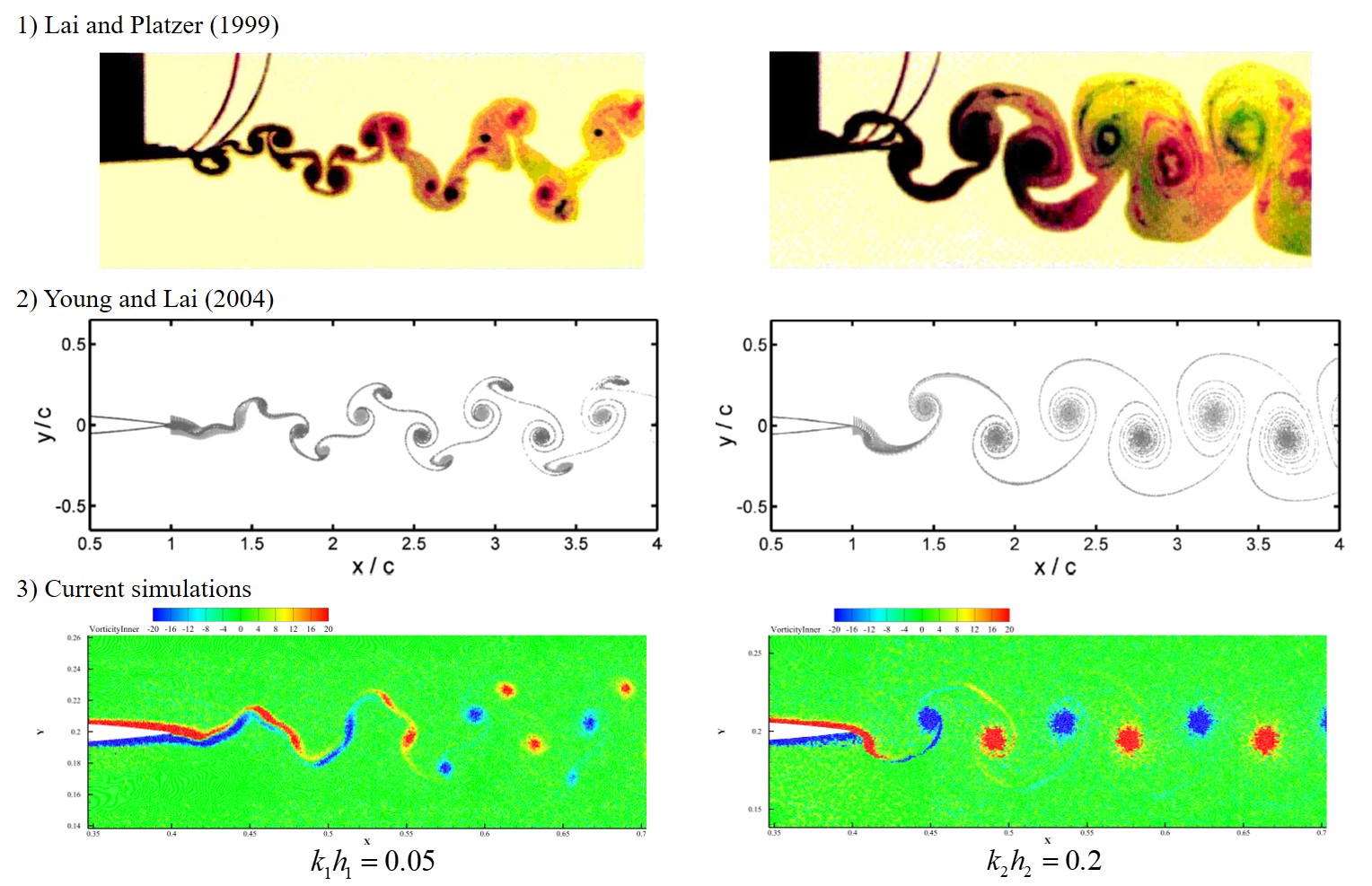}
\caption{Comparisons of vortex shedding patterns under different amplitudes and frequencies with experiments and simulations.}\label{fig27}
\end{figure}

\cite{lai1999jet} and \cite{young2004oscillation} conducted experimental and numerical studies, respectively, on a NACA0012 airfoil at \( Re = 20000 \). The results are shown in Figure \ref{fig27} with our simulation results with a particle resolution of $dp = 0.0004$ m. The reduced frequency is defined as \( k_1 = \pi f_1 L_1 / v_{in} \), and the dimensionless amplitude is given by \( h_1 = S_1 / L_1 \). We can clearly see that the vorticity field obtained from SPH simulations closely matches the experimental results, indicating that SPHinXsys can accurately describe the airfoil-like motion at this Reynolds number.

\subsubsection{Results}
\begin{figure}[htbp]
\centering
\includegraphics[width=\textwidth]{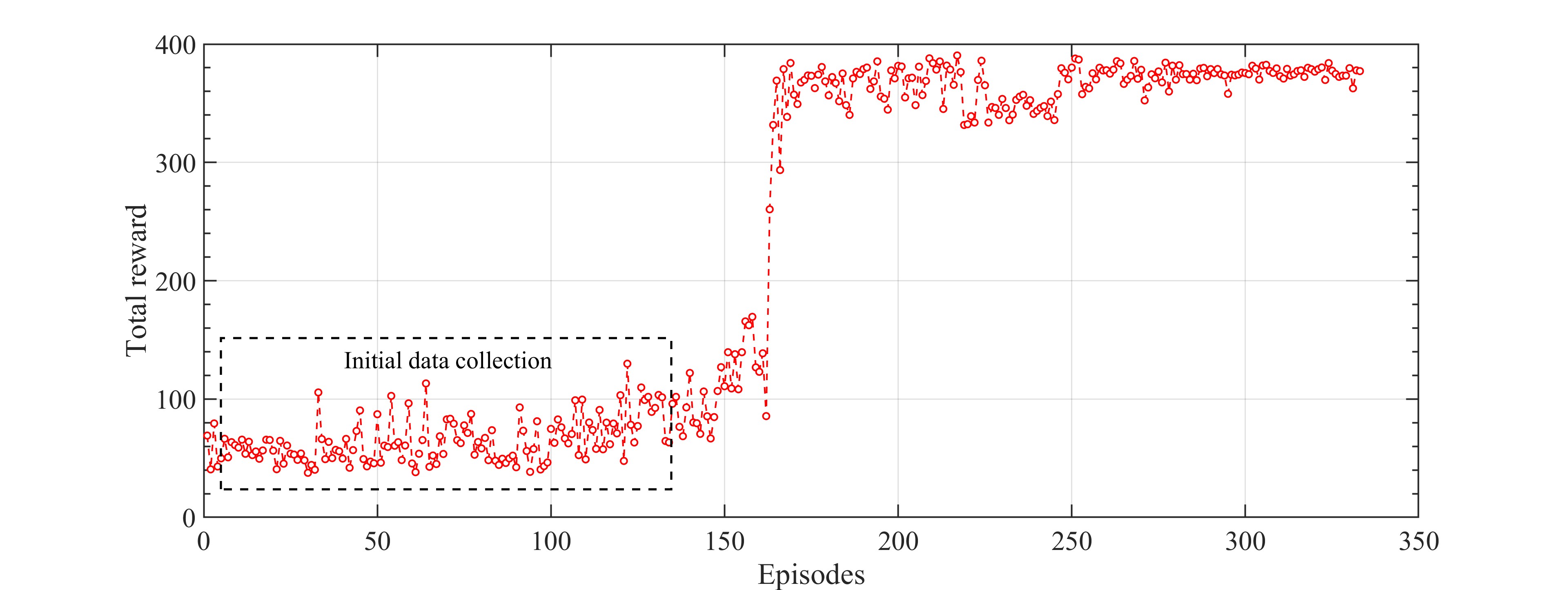}
\caption{The training curves obtained from training the muscle-driven fish to maintain a stable swimming position within the vortex street using the SAC algorithm.}\label{fig28}
\end{figure}

\begin{figure}[htbp]
\centering
\includegraphics[width=\textwidth]{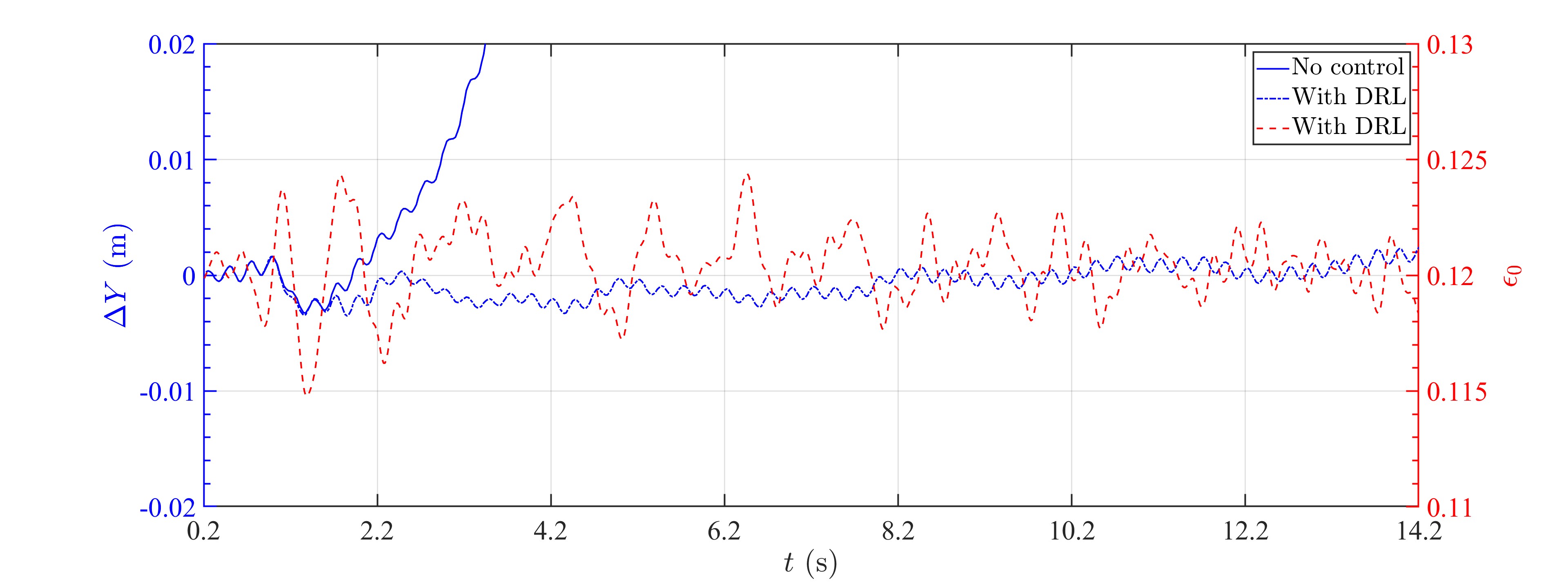}
\caption{The left blue curves represent the time variation of the fish's distance from the centerline within an episode, both with and without the DRL policy. The right curve shows the active strain variation of the red muscle on the fish's left side in the forward direction.}\label{fig29}
\end{figure}

Figure \ref{fig28} shows that after approximately 130 episodes of initial data collection, the agent discovers a relatively effective strategy around the 165th episode. The total reward per episode increases rapidly, and it converges to a stable policy after 250 episodes. As shown in Figure \ref{fig29}, the fish without control rapidly deviates from the centerline after 2.2 s and exits the computational domain around 4.2 s. In contrast, when the control policy is applied, the fish swims very stably along the centerline.

\begin{figure}[htbp]
\centering
\includegraphics[width=\textwidth]{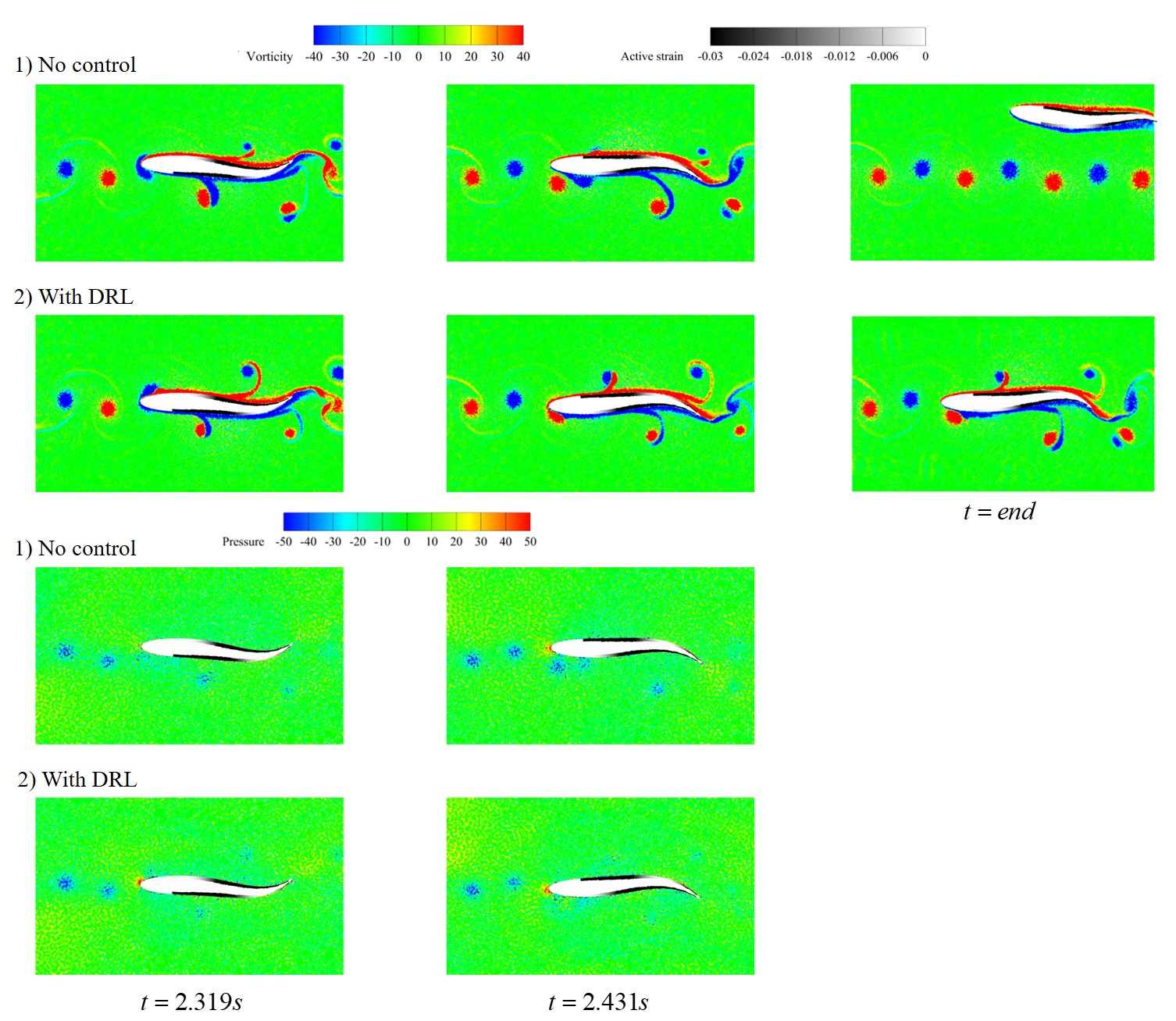}
\caption{The vorticity and pressure contour plots around the fish at three different time points, both with and without the control policy.}\label{fig30}
\end{figure}

From the contour plots in Figure \ref{fig30}, it is clear that the fish controlled by DRL can navigate through the gaps between two oppositely rotating vortices. The pressure distribution around the fish's head shows that as a vortex passes along the outer side of the head, a noticeable high-pressure region forms. This high-pressure region pushes the fish's head inward, allowing the fish to maintain a stable swimming pattern near the centerline. In contrast, at \( t = 2.319 \, \text{s} \), the fish, without a policy, fails to effectively utilize the vortices and is gradually pushed out of the central flow region. Besides, it is worth noting that once the fish is pushed out of the central flow region, its displacement in the x-direction changes significantly, gradually moving backward. However, the fish behind the airfoil can consistently swim near its initial x-position.

In conjunction with further analysis of the fish's swimming efficiency, we first define the net force on the fish \( F_{\text{net}} \) as \citep{curatolo2016modeling}
\begin{equation}
\label{eq:eq53}
\mathbf{F}_{net} = \sum_{a\in N}(\mathbf{f}_{ap} + \mathbf{f}_{a\nu}),
\end{equation}
then the thrust force $\mathbf{F}_{thrust}$ is given by
\begin{equation}
\label{eq:eq54}
\mathbf{F}_{thrust} = \frac{1}{2}(\mathbf{F}_{net} + \mathbf{F}_{abs}),
\end{equation}
where $\mathbf{F}_{abs} = \sum_{a\in N}|\mathbf{f}_{ap}| + \sum_{a\in N}|\mathbf{f}_{a\nu}|$. The work done by the active strain can be computed with
\begin{equation}
\label{eq:eq55}
W_a = \sum_{a\in N}\sigma_{active} \Delta\mathbb{E}_f V_a.
\end{equation}
Here $\sigma_{active} = \sigma_{total} - \sigma_{passive}$ is the active stress tensor, $V_a$ the fish particle volume. The swimming efficiency during a complete tail-beat cycle can be calculated with $\eta_f = (\sum_t \mathbf{F}_{thrust} \cdot \mathbf{v}_{fish}dt)/(\sum W_a)$. 

\begin{figure}[htbp]
\centering
\includegraphics[width=\textwidth]{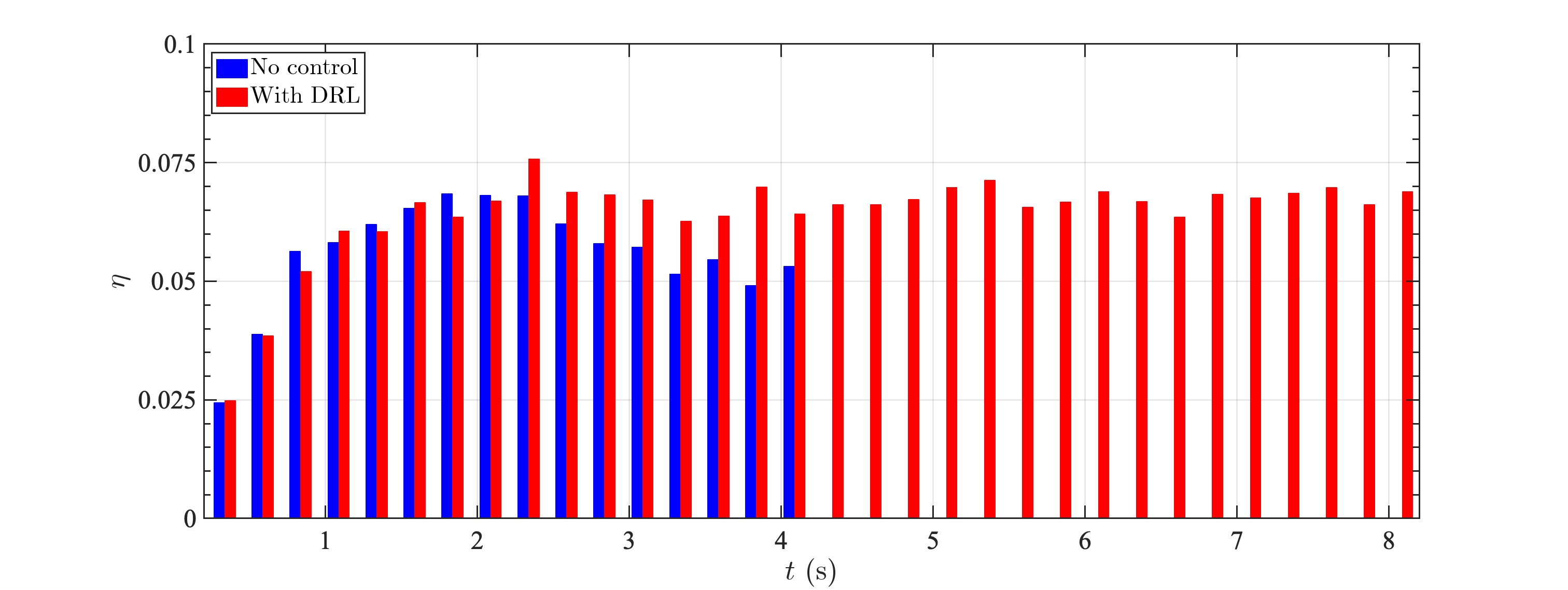}
\caption{The swimming efficiency of the fish over different tail-beat cycles, with and without control.}\label{fig31}
\end{figure}

As shown clearly in Figure \ref{fig31}, the swimming efficiency of the fish is significantly higher when moving within the vortex street compared to swimming in still water. Given that the total active strain is constant when controlling the fish's muscles, the active elastic energy applied in both swimming states can be considered equivalent. This indicates that the fish generates higher thrust in the vortex street, primarily due to its utilization of the low-pressure regions within the vortices. This finding is consistent with the experimental conclusions of \citep{thandiackal2023line}.

\section{Conclusion}
This study presents the development of DRLinSPH, an integrated platform designed to optimize AFC problems in FSI. The platform combines the self-programmed numerical software SPHinXsys, based on the SPH method, and the DRL platform Tianshou. On this foundation, four FSI-related case studies were constructed and optimized. 

The results of \emph{Case 1} and \emph{Case 2} demonstrate that, for sloshing suppression problems, applying additional displacement to the rigid baffle or active strain to the elastic baffle performs negative work on the liquid in the tank, thereby enhancing the suppression of sloshing at specific frequencies and amplitudes. Furthermore, smaller displacements or deformations of the baffle have a minimal impact on the sloshing frequency. 

\emph{Case 3} focusing on the OWSC, shows that optimizing the damping coefficient of the PTO system can directly alter the motion of the flap, increasing its deflection angle and improving wave energy capture, especially during the wave crest phase. 

\emph{Case 4} verifies that DRL can be used to control the red muscles of the fish, thus controlling its swimming direction and allowing it to follow a specific trajectory. Moreover, compared to a single fish swimming in water, the in-line swimming mode of fish can enhance swimming efficiency.

Besides, in all four cases, the SAC algorithm was successfully applied, illustrating the accuracy and flexibility of the SPH method for addressing FSI problems, as well as the broader potential of SAC in engineering applications compared to other DRL algorithms.

It is important to note that the current research is still limited to specific operating conditions. Further investigation is required to determine whether the strategies developed under these specific conditions can be generalized to other scenarios, particularly those with stronger nonlinearity, such as irregular waves. Moreover, the current optimization problems are restricted to 2D numerical simulations. Given that 3D simulations require substantial computational resources, directly using a 3D environment for DRL training is not feasible in the short term. Future research should explore whether the strategies trained in 2D can be effectively applied to 3D simulations. Additionally, the observations made in 2D can be regarded as incomplete observations in 3D. For RL problems involving incomplete observations, RNNs may offer a more effective solution than fully connected layers, and this approach warrants further exploration.

\section*{CRediT authorship contribution statement}
\textbf{M. Ye:} Validation, Software, Methodology, Investigation, Formal analysis, Writing - original draft, Writing - review \& editing, Visualization. \textbf{H. Ma:} Software, Writing – review \& editing, Investigation. \textbf{Y.R. Ren:} Resources, Formal analysis, Conceptualization. \textbf{C. Zhang:} Methodology, Data curation. \textbf{O.J. Haidn:} Validation, Supervision, Investigation. \textbf{X.Y. Hu:} Writing – review \& editing, Supervision, Methodology, Investigation, Conceptualization.

\section*{Declaration of interests}
The authors declare that they have no known competing financial interests or personal relationships that could have appeared to influence the work reported in this paper.

\section*{Acknowledgments}
Mai Ye is supported by the China Scholarship Council (No.202006120018) when he conducted this work. C. Zhang and X.Y. Hu would like to express their gratitude to Deutsche Forschungsgemeinschaft (DFG) for their sponsorship of this research under grant number DFG HU1527/12-4. The corresponding code of this work is available on GitHub at \url{https://github.com/Xiangyu-Hu/SPHinXsys}.

\bibliographystyle{elsarticle-harv} 
\bibliography{DRLinSPH}

\end{document}